\documentclass[english]{emulateapj}

\usepackage[T1]{fontenc}
\usepackage[latin9]{inputenc}
\usepackage{array}
\usepackage{float}	
\usepackage{amsmath}
\usepackage{color}

\makeatletter

\providecommand{\tabularnewline}{\\}

\@ifundefined{textcolor}{}
{%
 \definecolor{BLACK}{gray}{0}
 \definecolor{WHITE}{gray}{1}
 \definecolor{RED}{rgb}{1,0,0}
 \definecolor{GREEN}{rgb}{0,1,0}
 \definecolor{BLUE}{rgb}{0,0,1}
 \definecolor{CYAN}{cmyk}{1,0,0,0}
 \definecolor{MAGENTA}{cmyk}{0,1,0,0}
 \definecolor{YELLOW}{cmyk}{0,0,1,0}
 }

\makeatother

\usepackage[normalem]{ulem}  
\newcommand{\new}[1]{{\color[rgb]{0.8,0,0}#1}}

\newcommand{\old}[1]{{\color[rgb]{0.4,0.7,0.4}\sout{#1}}}
\newcommand{\blue}[1]{{\color[rgb]{0,0,1}#1}}

\usepackage{babel}

\usepackage{natbib}

\shorttitle{A Non-radial Oscillation in an AMXP?}
\shortauthors{Strohmayer \& Mahmoodifar}

\begin{document}

\title{A Non-radial Oscillation Mode in an Accreting Millisecond
Pulsar?}

\author{Tod Strohmayer$^1$ and Simin Mahmoodifar$^2$ \\ {\normalfont 
$^1$Astrophysics Science Division and Joint Space-Science Institute, NASA's 
Goddard Space Flight Center, Greenbelt, MD 20771, USA} \\ {\normalfont 
$^2$Department of Physics and Joint Space-Science Institute, University of 
Maryland College Park, MD 20742, USA}}


\begin{abstract}

We present results of targeted searches for signatures of non-radial
oscillation modes (such as r- and g-modes) in neutron stars using {\it
RXTE} data from several accreting millisecond X-ray pulsars
(AMXPs). We search for potentially coherent signals in the neutron
star rest frame by first removing the phase delays associated with the
star's binary motion and computing FFT power spectra of continuous
light curves with up to $2^{30}$ time bins. We search a range of
frequencies in which both r- and g-modes are theoretically expected to
reside. Using data from the discovery outburst of the 435 Hz pulsar
XTE J1751$-$305 we find a single candidate, coherent oscillation with
a frequency of $0.5727597 \times \nu_{spin} = 249.332609$ Hz, and a
fractional Fourier amplitude of $7.46 \times 10^{-4}$. We estimate the
significance of this feature at the $1.6 \times 10^{-3}$ level,
slightly better than a $3\sigma$ detection. Based on the
observed frequency we argue that possible mode identifications include
rotationally-modified g-modes associated with either a helium-rich
surface layer or a density discontinuity due to electron captures on
hydrogen in the accreted ocean. In the latter case the presence of
sufficient hydrogen in this ultracompact system with a likely
helium-rich donor would present an interesting puzzle.  Alternatively,
the frequency could be identified with that of an inertial mode or a core r-mode modified
by the presence of a solid crust, however, the r-mode amplitude required to account for the observed modulation amplitude would induce a large spin-down rate inconsistent with the observed pulse timing measurements.  For the AMXPs XTE J1814$-$338 and
NGC 6440 X-2 we do not find any candidate oscillation signals, and we
place upper limits on the fractional Fourier amplitude of any coherent
oscillations in our frequency search range of $7.8\times 10^{-4}$ and
$5.6 \times 10^{-3}$, respectively.  We briefly discuss the prospects
and sensitivity for similar searches with future, larger X-ray
collecting area missions.

\end{abstract}

\section{Introduction}

The study of global stellar oscillations can provide a powerful probe
of the interior properties of stars. A prime example of this is the
rich field of helioseismology.  By comparison, efforts to probe the
exotic interiors of neutron stars via similar methods are still in
their infancy, but recent observational results have provided new
impetus to further explore asteroseismology of neutron
stars. For example, observations of quasiperiodic oscillations (QPOs)
in the X-ray flux of highly magnetized neutron stars, ``magnetars''
\citep{1998ApJ...498L..45D, 2005ApJ...628L..53I, 2005ApJ...632L.111S,
2006ApJ...653..593S, 2006ApJ...637L.117W, 2006csxs.book..547W},
which have been linked to global torsional vibrations within the
star's crust, may ultimately provide a promising new probe of a
neutron star's internal composition and structure.  In addition to the
magnetar QPOs, burst oscillations, pulsations seen at or near the
neutron star spin frequency during thermonuclear X-ray bursts from
accreting, low mass X-ray binary (LMXB) neutron stars (see
\cite{2006csxs.book..113S,2012ARA&A..50..609W} for reviews on burst
oscillations), may also be linked to stellar pulsations. Although a
comprehensive understanding of the physics of these oscillations is
still being developed, one of the models that has been proposed to
explain them is the Rossby wave (r-mode) model,
which assumes that the oscillation is produced by a low-frequency
r-mode (Rossby wave) propagating in the neutron star surface
``ocean.''  In this case the r-mode modulates the temperature
distribution across the neutron star surface and the resulting angular
variations of the surface thermal emission--combined with the spin of
the star--produce pulsations in the X-ray flux observed from the
stellar surface. For example, \cite{2005MNRAS.361..659L} and
\cite{2005MNRAS.361..504H} have explored this model, and computed
light curves for small azimuthal wavenumber, $m$, surface r-modes on
rotating neutron stars.

Accretion-powered millisecond X-ray pulsars (AMXPs) also show
small-amplitude X-ray oscillations with periods equal to their spin
periods. To explain the low modulation amplitudes and nearly
sinusoidal waveforms in these sources, \cite{2009ApJ...705L..36L}
proposed a model in which the X-rays are emitted from a hot-spot that
is located at or near a magnetic pole of the star, and the magnetic
pole is assumed to be close to the spin axis of the star. When the
emitting region is close to the spin axis, a small variation in its
position can produce relatively large changes in the amplitude and
phase of the X-ray variations. \cite{2010MNRAS.405.1444L} and
\cite{2010MNRAS.409..481N} later suggested that global oscillations of
neutron stars (for example, r-modes) can periodically perturb such a
hot-spot and therefore the oscillation mode periods might potentially
be observable as X-ray flux oscillations from these sources (we
discuss this in more detail below).

The global oscillation spectrum of neutron stars is rich, and has been
classified according to the restoring force relevant to each
particular mode \citep{1988ApJ...325..725M}. For example, pressure
modes (p-modes and the f-mode) are primarily supported by internal
pressure fluctuations (essentially sound waves) in the star and have
frequencies in the $10$ kHz range that scale as $(\bar\rho)^{1/2}$,
where $\bar\rho$ is the stellar mean density. The successive overtones
of these modes have higher frequencies.  By overtones we mean modes
with an increasing number of nodes (zero crossings) in their radial
displacement eigenfunctions.  Gravity modes (g-modes) confined
primarily to the region above the solid crust have buoyancy as their
restoring force and frequencies in the $1-100$ Hz range (in the
slow-rotation limit). The overtones of these surface g-modes have
decreasing frequencies. The finite shear modulus of the neutron star
crust leads to additional, shear-dominated modes.  These include the
purely transverse torsional modes (t-modes), briefly mentioned above
in the context of magnetar QPOs, which have frequencies larger than
about $30$ Hz, and the s-modes, which possess both radial and
transverse displacements. For both classes of torsional modes the
overtones--whose radial eigenfunctions have at least one node in the
crust--can be thought of as shear waves traveling vertically through
the crust. They have frequencies in the kHz range that scale inversely
with the thickness of the neutron star crust.

In the case of rotating neutron stars another important class of
oscillations are the so-called inertial modes for
which the restoring force of the pulsations is provided by the
Coriolis force \citep{2000ApJ...529..997Y,
2000ApJS..129..353Y}. A well known sub-set of these are the 
r-modes that couple to gravitational radiation and can be driven
unstable by the Chandrasekhar-Friedman-Schutz (CFS) mechanism
\citep{1978ApJ...222..281F, 1998ApJ...502..708A,
1998ApJ...502..714F}.   Whether they are excited or not is a
competition between the driving due to the coupling to gravitational
radiation and the various mechanisms--such as bulk and shear
viscosity--that can damp the oscillations.  The damping and transport
properties, such as viscosity, heat conductivity and neutrino
emissivity, depend significantly on the phase of dense matter present
in the star, and since r-modes can both brake the star's rotation and
heat its interior, study of the spin and thermal evolution of neutron
stars can be a potentially important probe of the dense matter
interior \citep{2013ApJ...773..140M, 2012MNRAS.424...93H}.
Moreover, the co-rotating frame r-mode frequencies depend on the
stellar spin rate and the internal composition and structure of the
star \citep{1999PhRvD..60f4006L, 2000ApJS..129..353Y,
2012PhRvD..85b4007A}.  Thus, observations of the frequencies of
non-radial oscillation modes of neutron stars would be very useful in
probing their internal structure, but except for the magnetar QPOs
linked to crustal vibrations and perhaps the surface r-modes linked
with burst oscillations, there have been no other direct observations
of these oscillations.


It is relevant to ask the question of how the presence of non-radial
oscillations might be inferred from observations.
As noted briefly above in the context of burst oscillations, if an
r-mode modulates the temperature distribution across the neutron star
surface, then this may be revealed as a pulsation in the X-ray flux
from the star.  Another possibility is that surface motions induced by
a particular oscillation mode perturb the X-ray emitting hot-spot that
is present during the outbursts of accreting millisecond X-ray pulsars
(AMXPs). This mechanism seems most relevant for quasi-toroidal modes
(such as the r- and g-modes) in which the dominant motions are
transverse--locally parallel to the stellar surface--as opposed to
radial. Such transverse motions can deform an emitting region in a
periodic fashion and thus imprint the periodic deformation on the
observed light curve from the source.  Indeed,
\cite{2010MNRAS.409..481N} explored this mechanism, and computed the
resulting light curves from such a perturbed hot-spot on a rotating
neutron star.  Since the hot-spot rotates with the star it is
periodically deformed at the oscillation frequency of the mode as
measured in the co-rotating frame of the star.  They computed the
modulation that would be produced by a hot-spot that is perturbed by
the surface motions associated with a global r-mode, and showed that
the r-mode frequency specified in the co-rotating frame is imprinted
on the light curve seen by a distant observer.  We note that surface
g-modes also have dominant horizontal displacements and could also be
relevant in this context. Using this model they also demonstrated that
the observed modulation amplitude of the light curve could be used to
infer or constrain the mode amplitude.

In the limit of slow rotation it is well known that the r-mode
frequency in the co-rotating frame is given by $\omega = 2 m \Omega /
l (l+1)$, where $m$ and $l$ are the spherical harmonic indices that
describe the angular distribution of the dominant toroidal
displacement vector, and $\Omega$ is the stellar spin frequency. The
most unstable r-mode is that associated with $l = m = 2$, which has
the familiar frequency $\omega = 2\Omega /3$ in the co-rotating
frame. For more rapidly rotating neutron stars, like the AMXPs, the
r-mode frequency deviates from the above limit and is typically
calculated in an expansion in powers of the angular rotation frequency
(see, for example, \cite{1999ApJ...521..764L, 2000ApJS..129..353Y,
1998PhRvL..80.4843L, 2012PhRvD..85b4007A}). This leads to an
expression for $\omega$ of the form, $\omega = \Omega (\kappa_0 +
\kappa_2 \bar\Omega^2)$, where for the $l = m = 2$ r-mode, $\kappa_0 =
2/3$, $\bar\Omega^2 = \Omega^2(R^3/GM)$ and $\kappa_2$, which
represents the next-order correction to the r-mode frequency, depends
on the properties of the unperturbed stellar model, such as its
equation of state (EOS) and entropy stratification
\citep{2000ApJS..129..353Y, 2012PhRvD..85b4007A}. Thus, if an r-mode
frequency is detected it can potentially provide interesting
information about the stellar interior, and perhaps be used to
identify the dense matter phase present in the core (see, for example,
Figure 3 in Alford et al. 2012).

The above discussion ignores the effects that the solid crust of the
neutron star may have in modifying the r-modes and their surface
displacements.  For example, \cite{2001ApJ...546.1121Y} have
investigated the r-modes for neutron star models including a solid
crust and show that they are strongly influenced by mode coupling with
the crustal torsional modes (t-modes).  They found that this mode
coupling can reduce the r-mode frequency from $2\Omega /3$ to values
as low as $\Omega/2$ to $2\Omega /5$ (see their Figure 2), and the
reduction occurs at and above a critical rotational frequency that is
close to the fundamental torsional mode frequency (we discuss this in
more detail below).

Since the spin frequency of outbursting AMXPs can be tracked with high
precision, and the r-mode frequencies are computed as a series
expansion in powers of the spin frequency, it is possible to carry out
coherent, targeted searches in such sources for r-modes in a specific
range of frequencies both above and below its ``expected'' ($\Omega
\rightarrow 0$ limit) value, $2\Omega /3$. Similar arguments apply for
other modes as well, such as the surface g-modes, some of which have
frequencies that overlap the expected frequency range for the
r-modes. Here we present the results of power spectral searches for
the signatures of such modes using data from several AMXPs obtained
with the {\it Rossi X-ray Timing Explorer} (RXTE). It is not our
intent here to present an exhaustive search of all known AMXPs,
rather, we illustrate the methods and present results for three
sources; XTE J1751$-$305 (hereafter J1751), XTE J1814$-$338 (hereafter
J1814), and NGC 6440 X-2 (hereafter X-2), all of which are within the
nominal r-mode instability window computed for hadronic matter, and
which had the highest inferred r-mode amplitude upper limits in our
recent study \citep{2013ApJ...773..140M}.  We will present a study of
additional sources, including SAX J1808.4$-$3658, in a sequel.  The
paper is organized as follows.  In \S 2 we illustrate in some detail
our search analysis procedures using data from the 435 Hz AMXP J1751.
We also present the search results for this source and describe our
best detection candidate, which is at a frequency of $0.57276
\nu_{spin}$ (249.33 Hz).  In \S 3 we summarize our search results for
the additional targets, the 206 Hz pulsar X-2, and the 314 Hz pulsar
J1814. In \S 4 we discuss possible mode identifications for the best
candidate frequency in J1751.  We also briefly discuss how future
observations with larger collecting area missions, such as ESA's Large
Observatory for X-ray Timing (LOFT, \cite{2012SPIE.8443E..2DF}), and
the Advanced X-ray Timing Array (AXTAR, \cite{2010SPIE.7732E.134R})
can improve the sensitivity of such searches.  We conclude with a
brief summary of our findings in \S 5.

\section{A Coherent Search in XTE J1751$-$305}

The most sensitive search procedure for a particular timing signature,
such as a coherent pulsation, depends on the nature of that signature.
In the context of searches employing Fourier power spectra the
greatest sensitivity is achieved by matching the frequency resolution
of the power spectrum to the expected frequency bandwidth of the
signal.  Thus, for a highly coherent signal the greatest sensitivity
is achieved by maximizing the frequency resolution.  This effectively
means that one should compute a single Fourier power spectrum of the
longest time series obtainable from the available data.  While the
exact frequency bandwidth of a candidate signal is often not known
precisely, the work of \cite{2010MNRAS.409..481N} suggests that a
signal produced by perturbation of a hot-spot by an r-mode (or some
other non-radial mode) may be quite coherent.  On the other hand,
conditions in the neutron star surface layers can evolve as accretion
continues during an outburst and these sources are known to exhibit
timing noise that is likely associated with variations in the latitude
and azimuth of the accretion hot-spot
\citep{2009ApJ...698L..60P}, so such processes are likely to
limit the effective coherence of such signals.  Because of this, as
well as computational constraints, we restrict the size of the longest
light curves for Fourier analysis in this work to $N = 2^{30}$ time
bins.  For a sample rate of 2048 Hz this corresponds to a time
interval of 524,288 s, or about 6 days. Depending on the amount of
data present for a given source, one can then average several
independent power spectra and/or adjacent Fourier frequency bins to
search for signals with broader frequency bandwidths (such as
quasi-periodic oscillations, QPOs).

In order to carry out searches at the highest frequency resolution it
is necessary to remove as best as possible the frequency drifts
associated with the binary motion of the neutron star about the center
of mass of the system in which it resides. This effectively places the
observer at the center of mass of the binary system, a point from
which the neutron star is neither approaching nor receding.  These
considerations lead to the following basic steps we use to carry out a
search.  First, the X-ray event arrival times are corrected to the
Solar System barycenter. Next, we fit a model to the observed
orbit-induced phase variations. This orbit model is used to convert
each photon event arrival time to a neutron star rotation phase.
These phases are then converted back to fiducial times using the
best-determined spin frequency of the neutron star.  Finally, these
orbit corrected times can be used to compute a single light curve
which can then be Fourier analyzed using Fast Fourier Transform (FFT)
power spectral methods.

\begin{figure*}
\begin{center}
\includegraphics[scale=0.4]{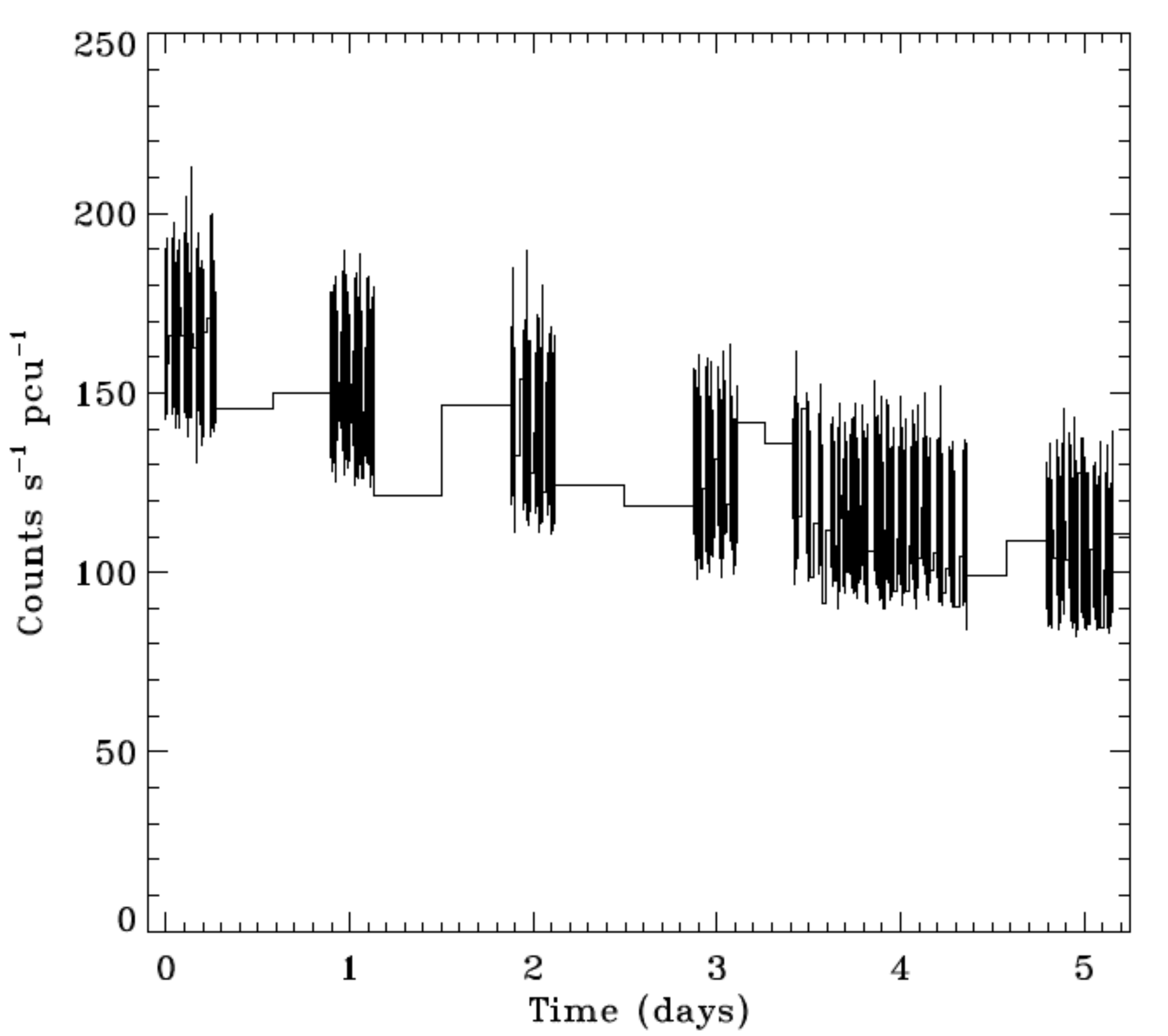}%
\end{center}
\caption{\label{fig:J1751-lc} Light curve in the 2 - 60 keV band from
XTE J1751-305 used in our pulsation search. These data span the
brightest portion of the outburst onset.  Time zero is 2002 Apr 05 at
15:29:03.422 UTC.  Note that the background level of $\approx 15$ 
counts s$^{-1}$ PCU$^{-1}$ has not been subtracted.}
\end{figure*}

\begin{figure*}
\begin{center}
\includegraphics[scale=0.4]{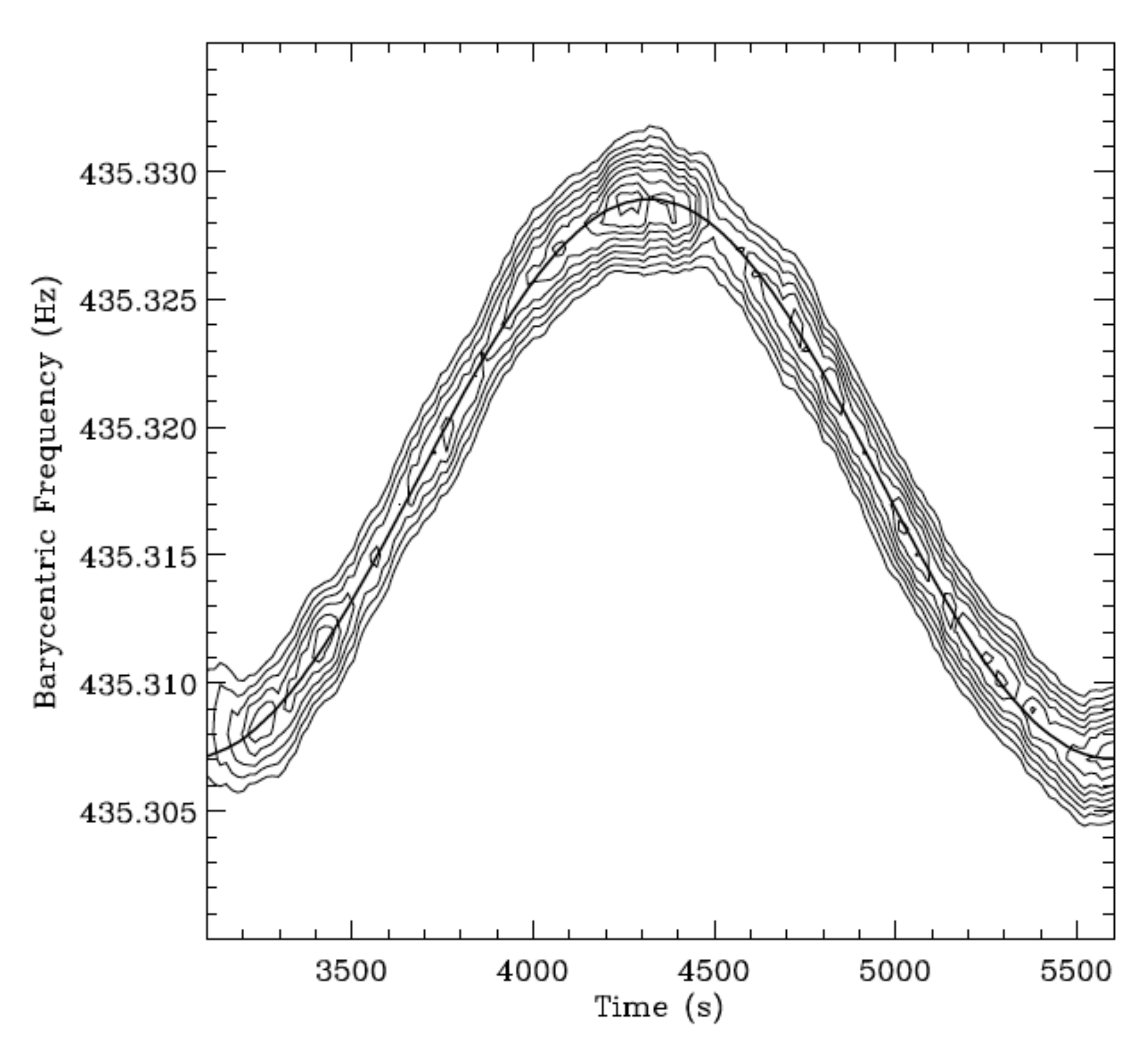}%
\end{center}
\caption{\label{fig:j1751_orbitmod} Dynamic power spectrum (Leahy-normalized) 
of XTE J1751-305 as a function of time and barycentric frequency in a single
{\it RXTE} orbit.  The contours show levels of Leahy-normalized
Fourier power and track the binary Doppler-shifted pulsar spin
frequency.  The solid curve is the best-fitting orbit model for this
data interval.  The origin for the time axis has the same reference as
Figure 1.}
\end{figure*}

\begin{figure*}
\begin{center}
\includegraphics[scale=0.4]{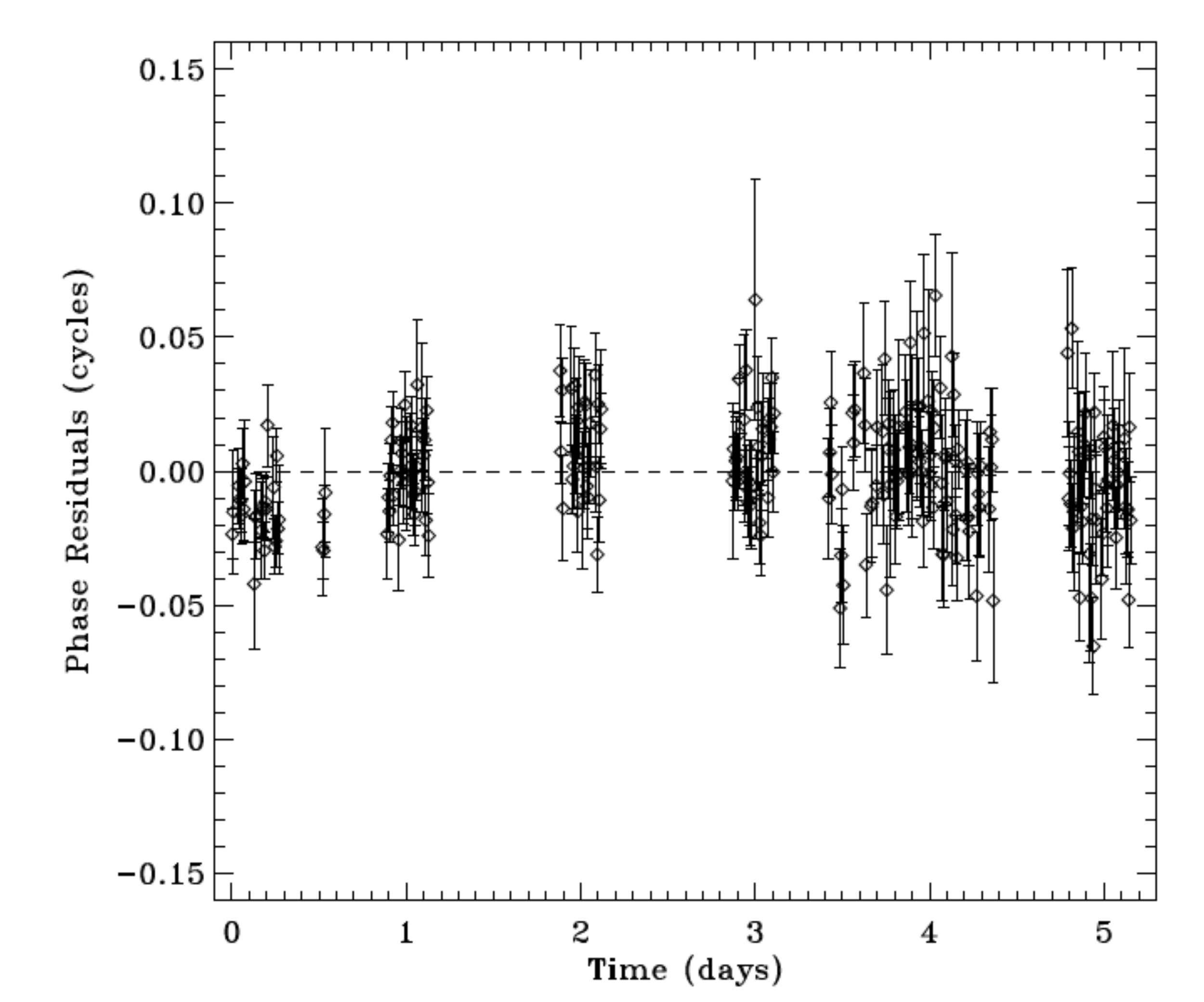}%
\end{center}
\caption{\label{fig:j1751_phaseres} Pulse timing phase residuals (in cycles)
for XTE J1751-305 after application of the best fitting circular orbit
model. The remaining phase residuals are poisson dominated. Time zero
is the same as in Figure 1.}
\end{figure*}

To illustrate the procedure in some detail we step through our
analysis for J1751.  This source was discovered in early April, 2002
during regular monitoring observations of the Galactic center region
using the RXTE Proportional Counter Array (PCA,
\cite{2002ApJ...575L..21M}). The outburst was relatively short,
lasting only about 10 days.  Timing of the X-ray pulsations revealed
an ultra-compact system with an orbital period of 42.4 min
\citep{2002ApJ...575L..21M}. For our coherent search we used data
spanning about 6 days during the peak of the
outburst. Figure~\ref{fig:J1751-lc} shows the source light curve
sampled in 2 s bins. We used PCA event mode data with a resolution of
125 $\mu$-sec for our study and included all events in the full energy
band-pass of the PCA and from all operating detectors. We used the
FTOOL {\it faxbary} to correct the photon arrival times to the Solar
System barycenter.  We then applied the orbit timing solution from
\cite{2002ApJ...575L..21M,2007ApJ...667L.211M} (see Table 1 in their
2007 paper) to convert the arrival times to neutron star rotational
phases.  Figure~\ref{fig:j1751_orbitmod} shows a dynamic power
spectrum from a single RXTE orbit, which reveals the time evolution of
the pulsar frequency due to the neutron star's orbital motion.  The
best fitting orbit model for this time interval (thick solid curve) is
also plotted, showing that it accurately predicts the observed
evolution. Figure~\ref{fig:j1751_phaseres} shows the resulting phase
residuals after application of the orbit model to the light curve used
for our coherent search. The remaining variations are consistent with
poisson errors in the phases.

\begin{figure*}
\begin{center}
\includegraphics[scale=0.4]{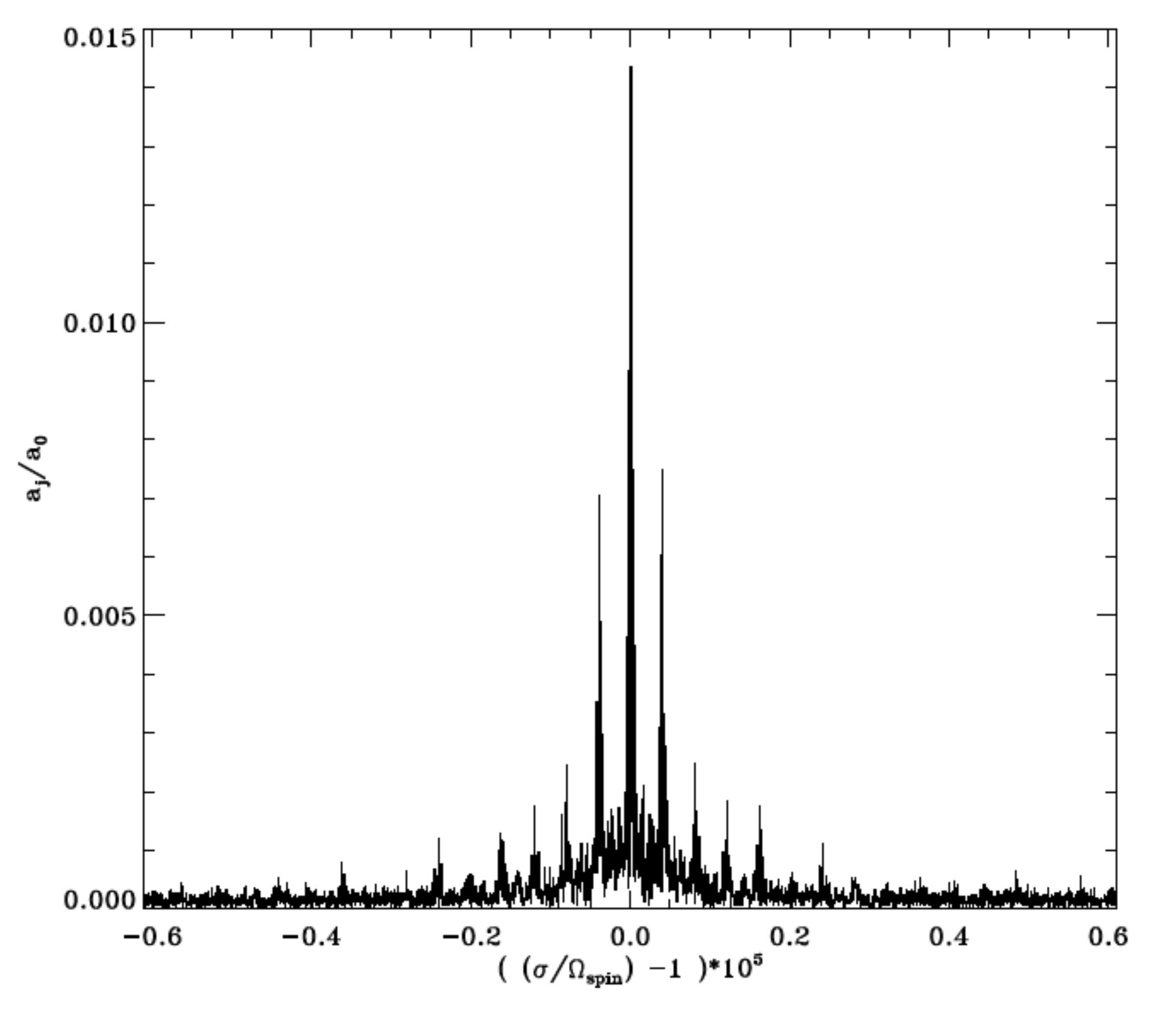}%
\end{center}
\caption{\label{fig:j1751_fundamental} A portion of the full frequency
resolution, coherent power spectrum for XTE J1751-305 in the vicinity
of the pulsar spin frequency (at 0 in these units). The power spectrum
is shown in units of Fourier amplitudes (see the text in \S 2 for
further details).  The side-lobe pattern of peaks results from the
uneven temporal sampling (the window function).}
\end{figure*}

We then used the orbit model to convert each arrival time to a
rotational phase. These phases can then be expressed as fiducial times
by multiplying by the best-fit pulsar spin period.  We use the
resulting times to produce a light curve sampled at 2048 Hz that
contains $2^{30} = 1,073,741,824$ time bins. Finally, we compute an
FFT power spectrum of this light curve. The resulting power spectrum
has a little more than half a billion frequency bins and a Nyquist
frequency of 1024 Hz, thus, simply from file size considerations it is
not practical to present a plot of the entire spectrum. However, to
demonstrate that the coherent pulsar signal is strongly detected we
show in Figure~\ref{fig:j1751_fundamental} the power spectrum in a
narrow frequency band centered on the pulsar signal.  Here, the units
on the x-axis are $(\sigma/\Omega - 1 )\times 10^5$, where $\sigma$ is
a Fourier frequency. Thus, the pulsar signal appears at zero in these
units. Moreover, in order to enable direct comparison with the light
curve computations of \cite{2010MNRAS.409..481N}, see for example
their Figure 6, we plot the power spectrum in units of fractional
Fourier amplitudes $\sqrt{(a_{j}^{*}a_{j})/N_{tot}}$, where the $a_j$
are the complex Fourier amplitudes at Fourier frequency $\nu_j = j /
(524,288 \; s)$, $j$ ranges from $0$ to $2^{29}$, $N_{tot}=
44,316,997$ is the total number of events in the light curve, and the
$*$ symbol indicates complex conjugation. To convert the fractional
Fourier amplitudes to the commonly used Leahy normalization one simply
squares the fractional amplitudes, and then multiplies by $2\times
N_{tot}$.  The commonly employed fractional rms amplitude is simply
$\sqrt{2}$ times the fractional Fourier amplitude defined above.

\subsection{Search for Co-rotating Frame Frequencies Consistent 
with r- and g-modes}

\begin{figure*}
\begin{center}
\includegraphics[scale=0.4]{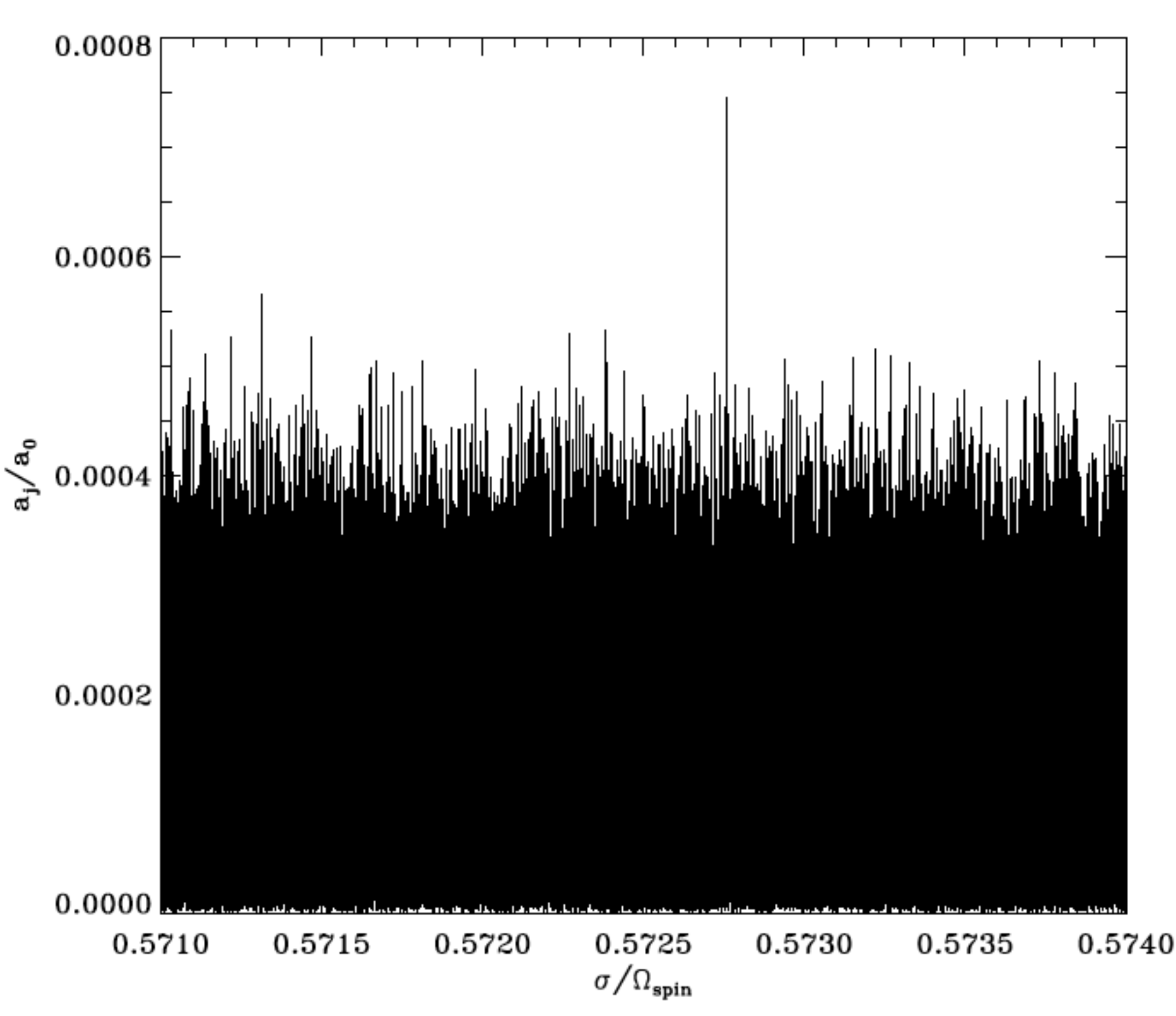}%
\end{center}
\caption{\label{fig:j1751_candidate} A portion of the full frequency
resolution, coherent power spectrum for XTE J1751-305 in the vicinity
of the candidate signal peak at $0.57276 \times \nu_{spin}$. The
spectrum is plotted in units of fractional Fourier amplitude.}
\end{figure*}

As noted in \S 1 above, when a pulsation mode periodically perturbs an
X-ray emitting hot-spot that is fixed in the rotating frame of the
star, the co-rotating frame mode frequency is imprinted on the light
curve seen by a distant observer. Further, rapid rotation tends to 
increase the co-rotating frame frequency
of the $l = m = 2$ r-mode from the slow-rotation limit of $\omega =
2\Omega /3$, while the influence of a solid crust may decrease it.
Based on the discussion above, a reasonable frequency range to search
is then $2/3 - k_1 \leq \omega/\Omega \leq (2/3 + k_2)$, where $k_1$
represents a plausible reduction in the frequency based on the
possible crustal effects to the r-mode, and $k_2$ represents a
reasonable maximum increase for $\kappa_2 \bar\Omega$ given the
observed spin frequency of J1751 and various possible masses,
equations of state and interior compositions for the neutron
star. Based on the calculations of \cite{2001ApJ...546.1121Y} and
Alford et al. (2012, see their Figure 3), plausible values for $k_1$
and $k_2$ are 0.25 and 0.09, respectively.  This defines a search
range from $0.4166 \leq \omega/\Omega \leq 0.75667$. A search in that
range reveals one candidate peak in slightly more than 77.59 million
independent Fourier frequency bins.  Figure~\ref{fig:j1751_candidate}
shows a portion of the full-resolution spectrum in the vicinity of
this peak. It appears at a frequency of $0.5727597 \times \nu_{spin} =
249.332609$ Hz, and has a fractional Fourier amplitude of $7.455
\times 10^{-4}$.

\begin{figure*}
\begin{center}
\includegraphics[scale=0.4]{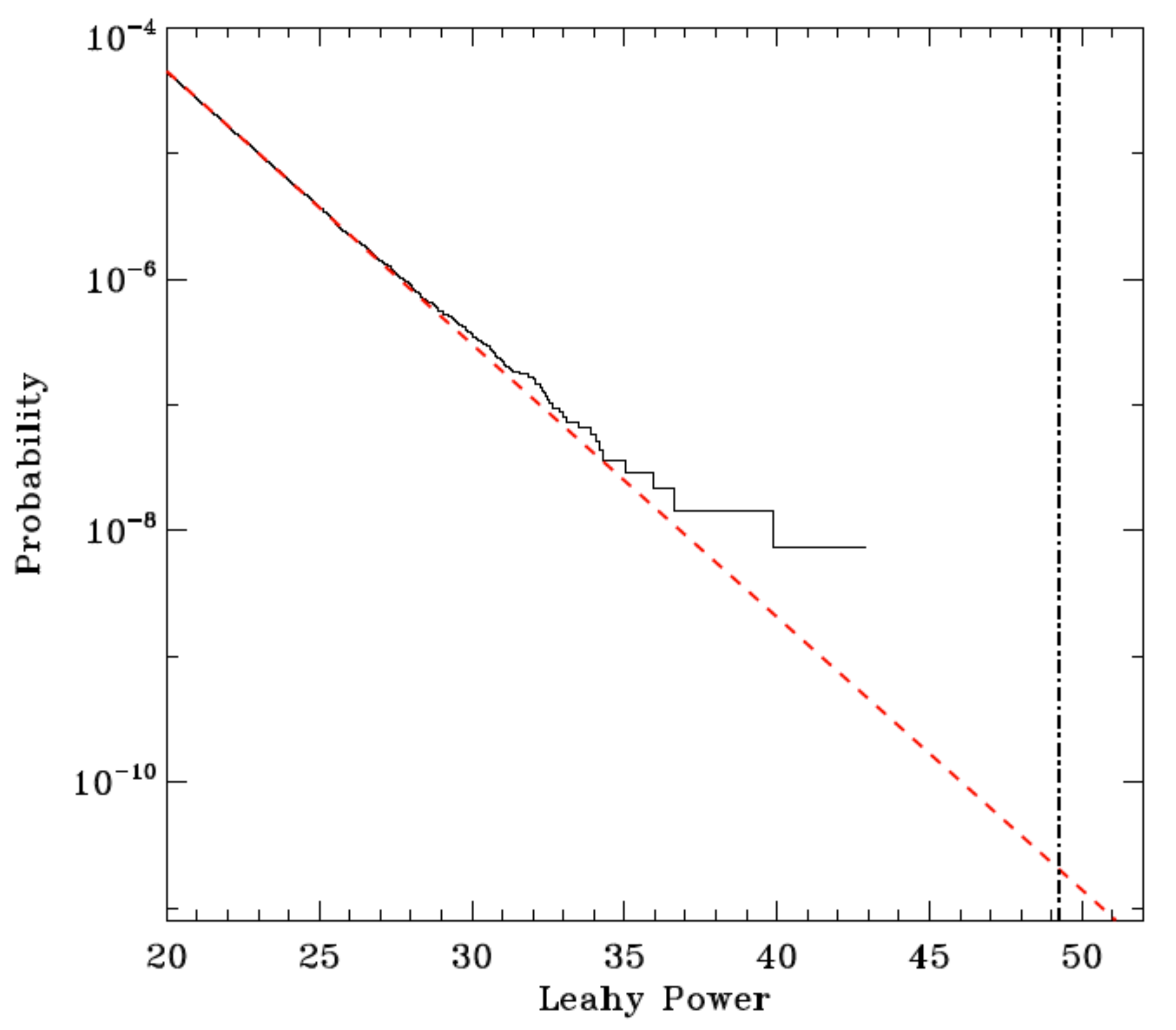}%
\end{center}
\caption{\label{fig:j1751_probtoexceed} Probability to exceed a given
Leahy-normalized Fourier power in a single trial.  The red squares
show the expected noise-power distribution (in this case a $\chi^2$
distribution with 2 degrees of freedom). The solid histogram shows the
observed power-spectral distribution for XTE J1751-305 in the frequency
range from 1.6 to 2.2 $\times$ the pulsar spin frequency.  The
vertical dashed line marks the power value of the candidate signal
peak.  The data track the expected distribution over a broad range of
power values.  An exact match at the highest power values is not
expected simply due to statistical fluctuations.}
\end{figure*}

To assess the significance of this peak we first convert its
fractional Fourier amplitude to a Leahy-normalized power and then
estimate its single-trial probability using the expected noise power
distribution, which for a single power spectrum is the $\chi^2$
distribution with 2 degrees of freedom. The peak Leahy-normalized
power is then 49.26, which corresponds to a single-trial probability
of $2\times 10^{-11}$.  Accounting for the number of trials by
multiplying by the number of independent Fourier frequencies in the
search range, $77.6 \times 10^{6}$, gives a significance of $1.6
\times 10^{-3}$, which is a little better than a $3\sigma$
detection. We then used a portion of the power spectrum at higher
frequencies (from 1.6 to 2.2 times the pulsar spin frequency) to
investigate how accurately the distribution of noise powers follows
the expected $\chi^2$ distribution. The result is shown in
Figure~\ref{fig:j1751_probtoexceed}, where the red dashed line denotes
the probability to exceed a given Fourier power for the $\chi^2$
distribution with 2 degrees of freedom, and the Leahy-normalized power
spectral data are plotted as a histogram.  Over the range of Fourier
powers present in the data the power spectral values show a good match
to the expected distribution. The Fourier power of the candidate peak
is marked by the vertical dashed-dot line, and as indicated above, has
a single trial probability of $2\times 10^{-11}$. Based on this we
think our significance estimate is reasonable.

\begin{figure*}
\begin{center}
\includegraphics[scale=0.4]{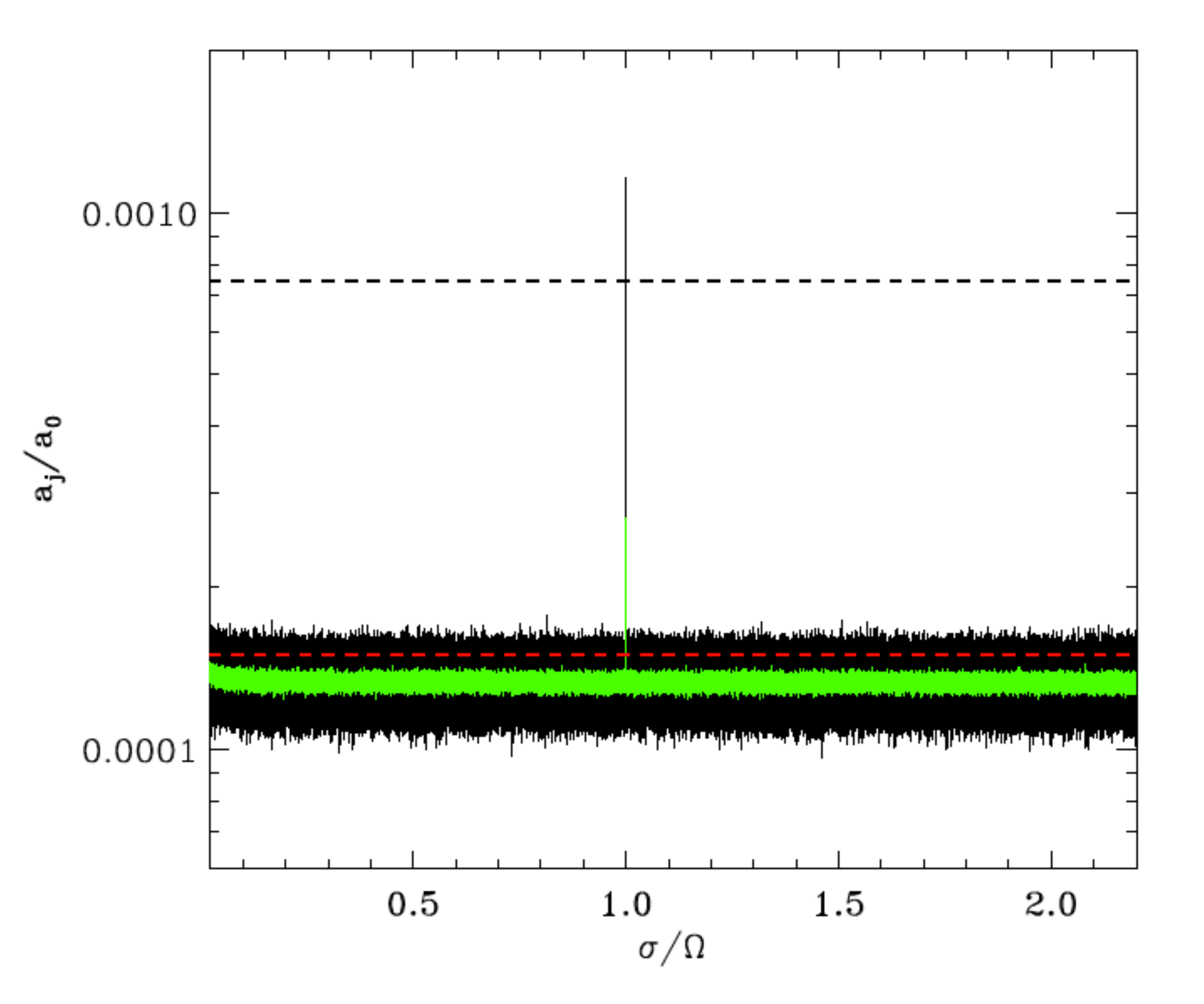}%
\end{center}
\caption{\label{fig:j1751_psd} Frequency-averaged power spectra of 
XTE J1751-305 plotted in units of fractional Fourier amplitude.  Spectra
averaged to 1/2048 (black) and 1/128 (green) Hz are shown. The X-axis
shows frequency scaled by the pulsar spin frequency. The pulsar signal
at 1 is clearly evident, but there are no other significant features
evident at either resolution. The horizontal dashed red line marks the
amplitude given by $1/\sqrt{N_{tot}}$, where $N_{tot}$ is the total
number of counts in the light curve.  The horizontal dashed line marks
the amplitude of the candidate signal peak at $0.57276 \times
\nu_{spin}$.}
\end{figure*}

We next averaged the full resolution power spectrum in order to search
for any broader bandwidth signals that might be present.
Figure~\ref{fig:j1751_psd} shows two such averaged power spectra over
the full frequency range. The black and green histograms have
frequency resolutions of 1/2048, and 1/128 Hz, respectively.  The
pulsar signal is still easily detected in each case, but we do not
find any other significant features at these or other frequency
resolutions. The horizontal dashed line marks the amplitude of the
candidate signal at $249.33$ Hz discussed above. We can place upper
limits on any signal power in our defined search range at these
frequency resolutions of $1.64 \times 10^{-4}$ and $1.42 \times
10^{-4}$, respectively.  The horizontal, red dashed line in
Figure~\ref{fig:j1751_psd} marks an amplitude given by
$1/(N_{tot})^{1/2} = 1.50 \times 10^{-4}$, which gives a reasonably
close approximation to the quoted upper limits for broader band
signals.

\subsection{Search for Modulation at the Inertial Frame r-mode Frequency}

As discussed in \S 1, if an oscillation mode modulates emission
over the entire neutron star surface rather than simply
perturbing a hot-spot fixed in the co-rotating frame, then one would
expect a pulsation signal at the mode's inertial frame
frequency, $\omega_i = 2\Omega - \omega$, where $\omega$ is the
co-rotating frame frequency.  Thus, to search the range of inertial
frame frequencies corresponding to the range of co-rotating frame
frequencies just discussed in \S 2.1 we need to search the
frequency range $2 - (2/3 + 0.09) < \sigma / \Omega < 2 - (2/3 -
0.25)$, which reduces to $1.243 < \sigma / \Omega < 1.583$.  A search
reveals no significant peaks in this range. The highest peak appears
at a frequency of $1.565327 \nu_{spin}$, with a fractional Fourier
amplitude limit of $6.6\times 10^{-4}$.

\section{Coherent Searches in XTE J1814$-$338 and NGC 6440 X-2}

Here we briefly summarize search results for J1814 and X-2.

\subsection{Results for XTE J1814$-$338}

\begin{figure*}
\begin{center}
\includegraphics[scale=0.4]{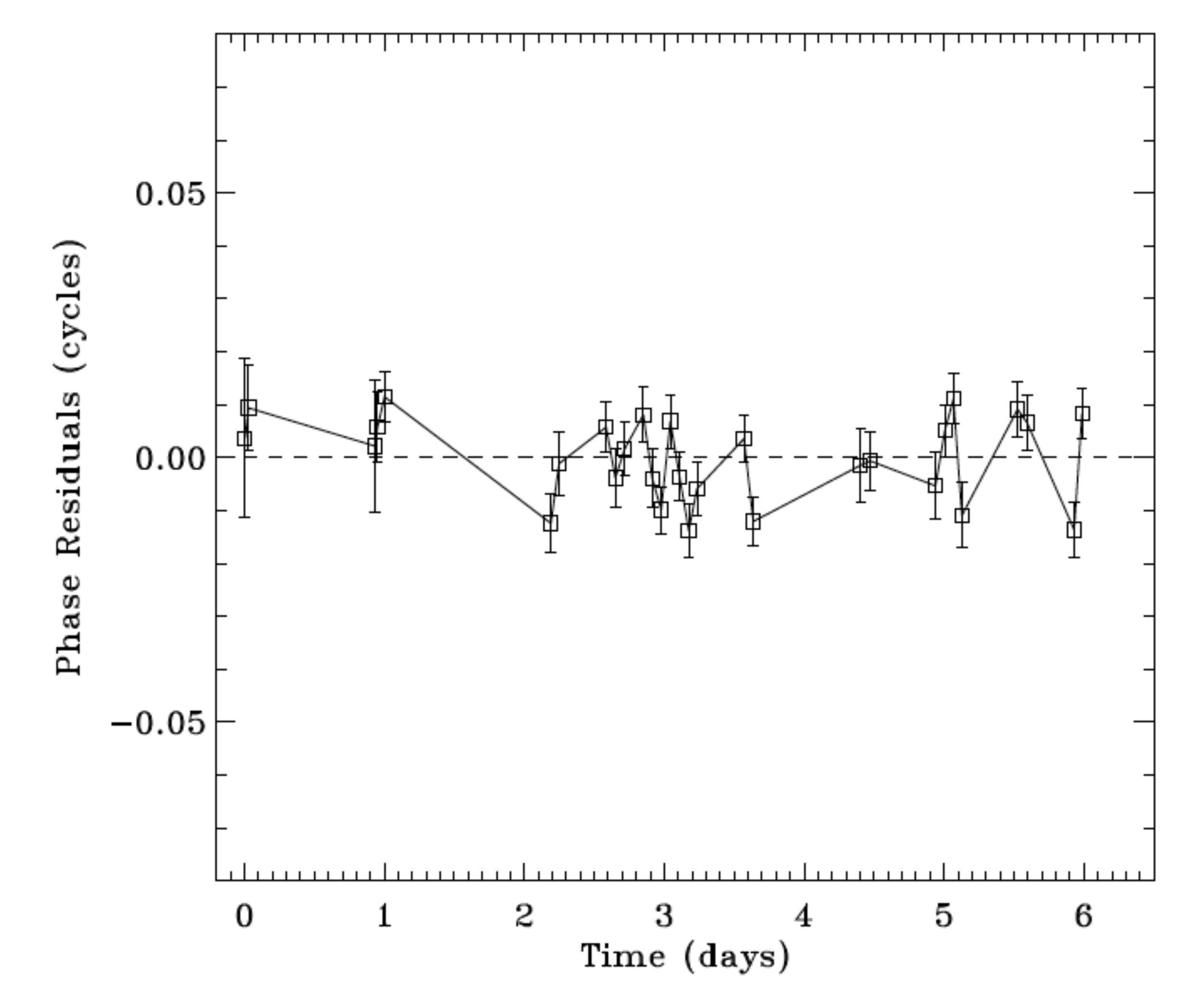}%
\end{center}
\caption{\label{fig:j1814_phaseres1} Pulse timing phase residuals (in cycles)
for XTE J1814-338 (first time interval) after application of the best
fitting circular orbit model. Time zero is 2003 June 5 at 02:34:20 UTC.} 
%
\end{figure*}

\begin{figure*}
\begin{center}
\includegraphics[scale=0.4]{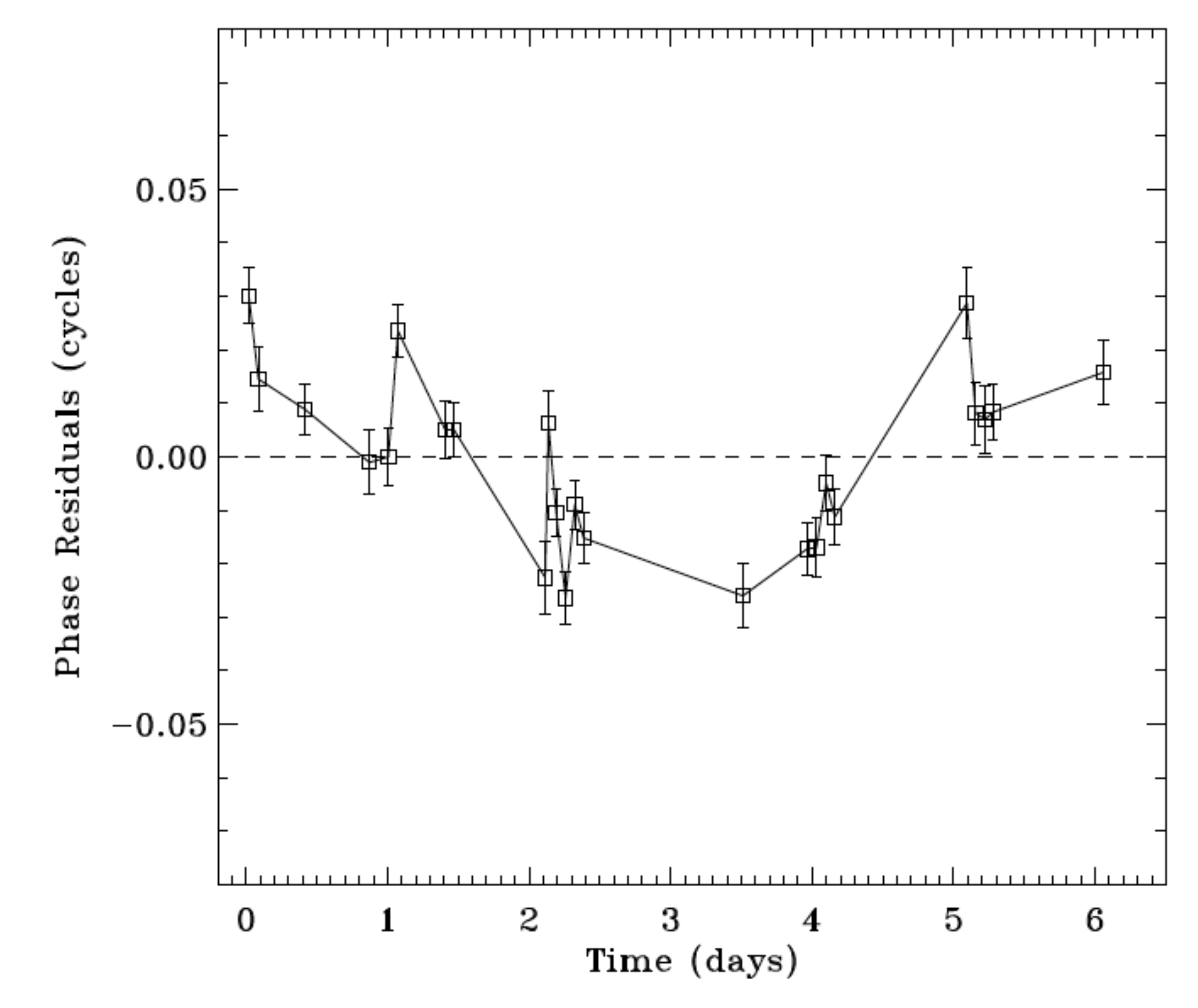}%
\end{center}
\caption{\label{fig:j1814_phaseres2} Pulse timing phase residuals (in cycles)
for XTE J1814-338 (second time interval) after application of the best
fitting circular orbit model. Time zero is 2003 June 11 at 04:12:28 UTC.}
%
\end{figure*}

\begin{figure*}
\begin{center}
\includegraphics[scale=0.4]{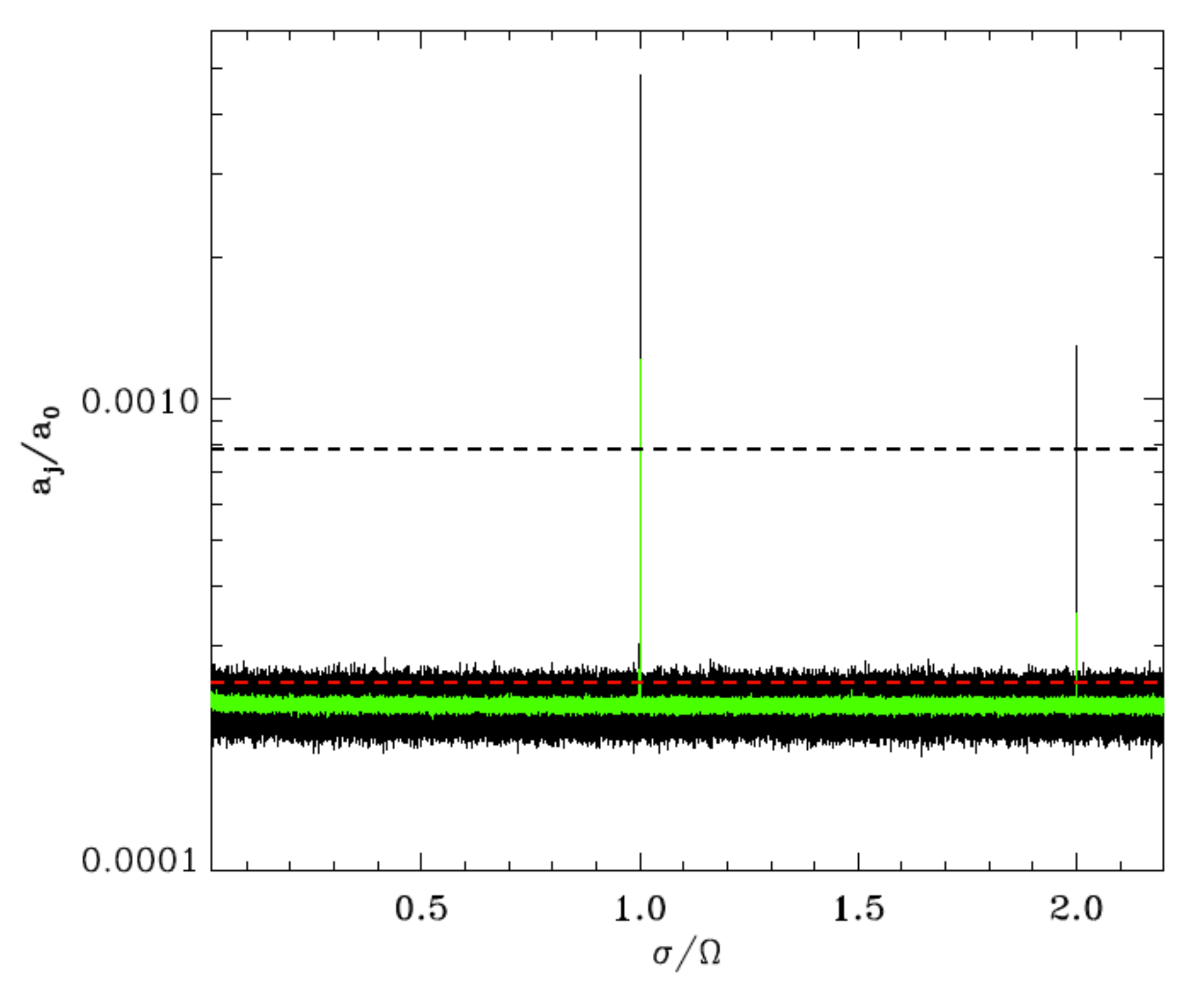}%
\end{center}
\caption{\label{fig:j1814_psd} Frequency-averaged power spectra of
XTE J1814-338 (average of both data intervals analyzed) plotted in units of
fractional Fourier amplitude.  Spectra averaged to 1/2048 (black) and
1/128 (green) Hz are shown. The x-axis shows frequency scaled by the
pulsar spin frequency. The pulsar fundamental and first harmonic are
clearly evident, but there are no other significant features evident
at either resolution. The horizontal dashed red line marks the
amplitude given by $1/\sqrt{N_{tot}/2}$, where $N_{tot}$ is the total
number of counts in the light curve.  The horizontal dashed line
(black) marks the upper limit on the amplitude at the full frequency
resolution.}

\end{figure*}

J1814 was discovered by {\it RXTE} in June 2003 using data obtained
with the Galactic bulge monitoring program then being conducted with
the PCA onboard {\it RXTE}.  The pulsar has a 314.36 Hz spin frequency
and an orbital period of 4.275 hr
\citep{2003ATel..164....1M,2007MNRAS.375..971P}. The discovery
outburst lasted for $\approx 50$ days. This object was the first
neutron star to exhibit burst oscillations with a significant first
harmonic \citep{2003ApJ...596L..67S}, and indeed, the persistent pulse
profile also shows substantial harmonic content.  This source is also
known to exhibit significant timing noise, that is, systematic timing
residuals remain after modeling the binary Doppler delays
\citep{2007MNRAS.375..971P, 2008ApJ...688L..37W}.  This noise is still
not completely understood, but may represent movement of the accretion
hot-spot relative to the stellar spin axis as the accretion rate
changes during an outburst \citep{2010ApJ...722..909P}. Here we use
data from the first 12 days for which such variations were less
significant \citep{2008ApJ...688L..37W}.  We used data beginning on
June 5, 2003 at 02:34:20 UTC and constructed two light curves, each
sampled at 2048 Hz and with $2^{30}$ time bins. There are a total of
$31,361,962$ X-ray events in the two light curves. We first modeled
the orbital variations in a similar manner as described above for
J1751.  Our orbit parameters are consistent with those of
\cite{2007MNRAS.375..971P}. Figures~\ref{fig:j1814_phaseres1} and
\ref{fig:j1814_phaseres2} show the resulting orbit-corrected phase
residuals for the two data segments used to construct our light
curves.  One can see that the second interval
(Figure~\ref{fig:j1814_phaseres2}) shows more systematic timing noise
than the first interval (Figure~\ref{fig:j1814_phaseres1}).  We then
computed power spectra for each interval in the same manner as
described for J1751. We searched the power spectra in the same
frequency ranges as described above for J1751 and for each data
interval separately as well as the average power spectrum computed
from both intervals.  We did not find any significant features in the
power spectra.  Figure~\ref{fig:j1814_psd} shows two average power
spectra computed from both intervals, the black and green histograms
have been averaged to frequency resolutions of 1/2048, and 1/128 Hz,
respectively.  The pulsar fundamental and first harmonic are clearly
evident (at 1 and 2 in these units). The horizontal dashed line
(black) marks the upper limit of $7.8\times 10^{-4}$ on any signal
power at the full frequency resolution of the power spectrum.  The
horizontal dashed (red) line marks an amplitude given by
$1/(N_{tot}/2)^{1/2} \approx 2.5 \times 10^{-4}$, which again gives a
reasonably close approximation to the upper limits for broader band
signals.

\subsection{Results for NGC 6440 X-2}

\begin{figure*}
\begin{center}
\includegraphics[scale=0.4]{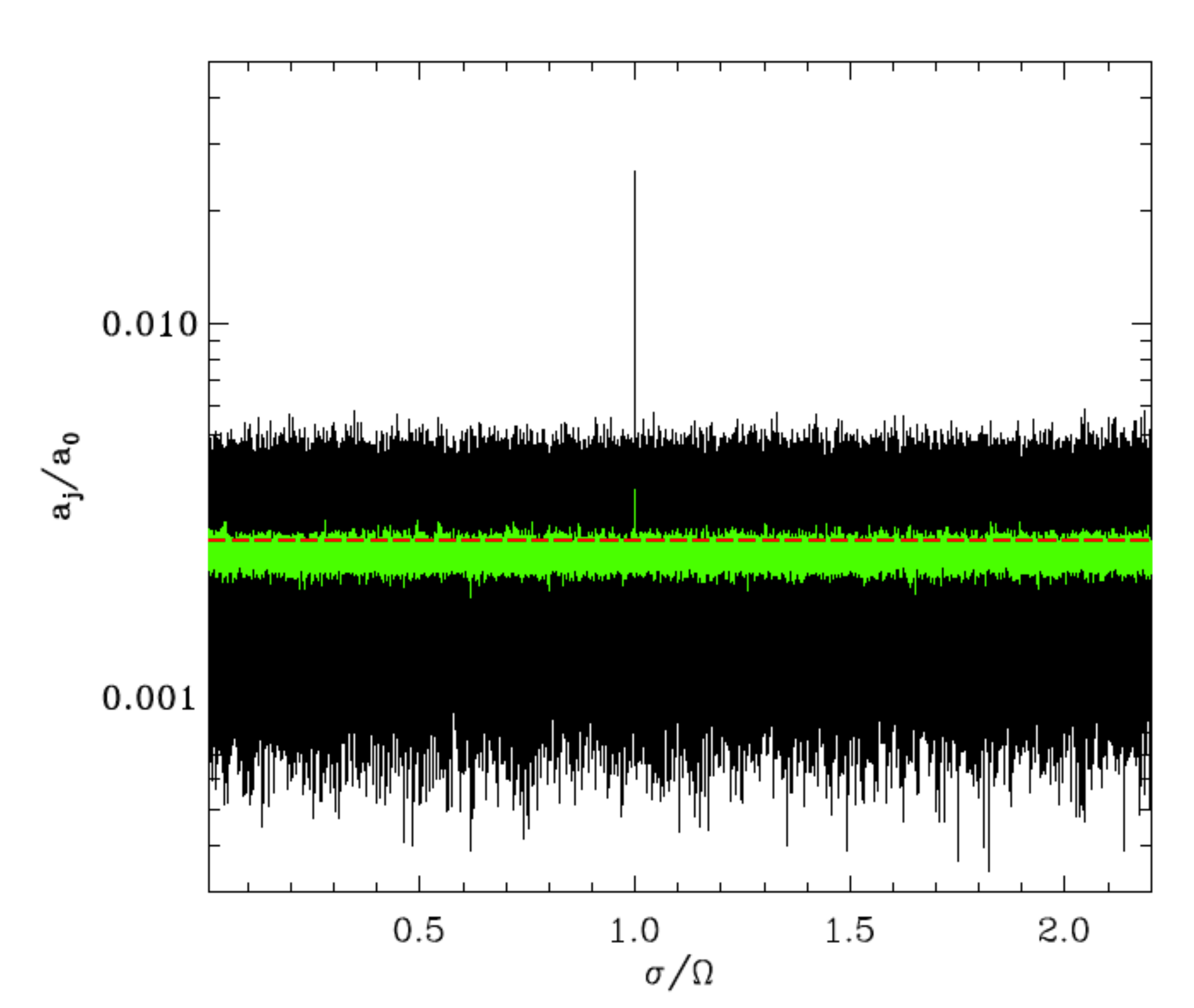}%
\end{center}
\caption{\label{fig:ngc_psd} Frequency-averaged power spectra of NGC 6440 X-2
plotted in units of fractional Fourier amplitude.  Spectra at the full
frequency resolution (1/3200 Hz, black) and 0.01 Hz (red) are
shown. The x-axis shows frequency scaled by the pulsar spin
frequency. The pulsar signal at 1 is clearly evident, but there are no
other significant features evident at either resolution. The
horizontal dashed red line marks the amplitude given by
$1/\sqrt{N_{tot}/4}$, where $N_{tot}$ is the total number of counts in
the four light curves analyzed.}
\end{figure*}

Pulsations at 205.89 Hz were detected with {\it RXTE} from the
globular cluster source NGC 6440 X-2 on 30 August, 2009
\citep{2009ATel.2182....1A}.  On this date the source was observed for
a single {\it RXTE} orbit, yielding $\approx 3000$ s of exposure,
revealing a 57 min orbital period \citep{2010ApJ...712L..58A}.  A
subsequent outburst with detectable pulsations was observed with {\it
RXTE} on 21 March, 2010, for an additional 3 {\it RXTE} orbits and
6600 s of exposure.  We used all these data in our search.  As for
J1751, we first barycentered the data using the best determined
position from \citet{2010ApJ...714..894H}. Because the available data
for this source are too sparse to enable calculation of a single,
coherent Fourier power spectrum, we separately modeled the orbital
variations in each of the four data segments. We then generated light
curves using the orbit-corrected arrival times, computed a Fourier
power spectrum for each, and then averaged them. The light curves were
sampled at 8192 Hz, yielding a Nyquist frequency of 4096 Hz. The
resulting averaged power spectrum is shown in
Figure~\ref{fig:ngc_psd}.  Since there are many fewer Fourier
frequencies compared to either J1751 or J1814, we show the spectrum at
the full frequency resolution.
Two spectra are shown in Figure~\ref{fig:ngc_psd}, the black curve is
plotted at the full frequency resolution ($3.125 \times 10^{-4}$ Hz),
and the green curve has been averaged by a factor of 32 to a
resolution of 0.01 Hz. The pulsar fundamental is clearly evident (at 1
in these units), but there are no other candidate detections. At these
resolutions we can place upper limits on any signal power in our
search ranges of $5.6 \times 10^{-3}$ (at $3.125 \times 10^{-4}$ Hz
resolution), and $2.8 \times 10^{-3}$ (at $0.02$ Hz). Because much
less data is available for X-2 than for either J1751 or J1814, the
limits are not as constraining as for those sources.

\section{Discussion} 

As discussed in \S 2, we found a candidate oscillation at a frequency
$\omega =0.5727 \Omega$ with an estimated significance of $1.6 \times
10^{-3}$ in data from the discovery outburst of J$1751$. Here we
discuss possible mode identifications for this candidate oscillation.
As mentioned in the introduction, AMXPs show small-amplitude X-ray
oscillations with periods equal to the spin period of the star. To
explain their low modulation amplitudes and nearly sinusoidal waveforms
\cite{2009ApJ...705L..36L} proposed a model in which
X-rays are emitted from a hot-spot at the stellar surface and near a
magnetic pole that is assumed to be close to the
rotation axis of the star. If we assume that this model is correct,
then transverse motions induced by the non-radial oscillations at the
surface of the star can perturb the hot-spot periodically, and these
periodicities might be observable in the radiation flux from the star
\citep{2010MNRAS.409..481N}.  In addition to producing X-ray
variations by perturbing the hot-spot, if the amplitude of the
oscillations at the surface of the star are large enough they might
also generate X-ray variations by modulating the surface temperature
of the star \citep{2005MNRAS.361..659L}. In the former case--where the
surface oscillation perturbs the hot-spot--since it is co-moving with
the star, a distant observer will detect the oscillation frequency of
the mode as measured in the co-rotating frame of the star (we refer to
this as the ``co-rotating frame scenario''). This has been shown by
\cite{2010MNRAS.409..481N} for the case of r-modes.  On the other
hand, oscillation-induced temperature perturbations will produce X-ray
variations with the same periodicity as the oscillation frequency of
the mode as measured in an inertial frame (we call this the ``inertial
frame scenario'').

For fast rotating, accreting neutron stars such as J1751 the most
relevant restoring forces that can produce stellar pulsations with
frequencies that might be consistent with the candidate frequency in
J1751 are the coriolis force--due to the star's rotation--and buoyancy
associated with thermal and composition gradients.  As briefly summarized 
in \S 1, the corresponding oscillation
modes associated with these forces are the inertial modes (which
includes the r-modes) and the gravity modes (g-modes).  In general,
both forces are present and the nature of the resulting modes will
depend on their relative strength. For example, at high rotation rates
the coriolis force will almost certainly dominate--except perhaps
within a very small band around the rotational equator--and the
resulting pulsation modes are expected to be inertial in character. At
the other extreme of slow rotation buoyancy can eventually prevail
resulting in essentially pure g-modes \citep{2000ApJS..129..353Y,
2009MNRAS.394..730P}. Other oscillation modes such as crustal toroidal
modes associated with the finite shear modulus of the crust, or f- and
p-modes due to pressure forces are either confined to the crust or core 
of the star and may not be able to induce motions at the
surface, or they have higher frequencies which are inconsistent with
the candidate oscillation.  In what follows we discuss how the
candidate frequency in J$1751$ might be identified as a surface
g-mode, a core r-mode, or perhaps an inertial mode in a fast rotating
star.


\subsection{g-modes}
As briefly mentioned above, the g-modes are low frequency non-radial
oscillation modes of neutron stars with buoyancy as their restoring
force. In a 3 component NS model composed of a fluid core, a solid
crust and a fluid ocean, g-modes might be excited in the core and/or
in the ocean, but the finite shear modulus excludes them from the
crust \citep{1995ApJ...449..800B}. As a result core g-modes are
unlikely to have observable effects on the radiation observed from the
surface of the star. Therefore, here we focus on the surface g-modes
that are confined to a thin layer at the surface of the star. There
have been many studies on g-modes in neutron stars
\citep{1987ApJ...318..278M, 1996ApJ...467..773S, 1995ApJ...449..800B,
1996ApJ...460..827B, 1998ApJ...506..842B}. We are particularly
interested in the surface g-modes in AMXPs. The conditions at the
surface of these objects evolve slowly due to the accretion and their
g-mode spectrum is different from that of the isolated and
non-accreting neutron stars. The g-modes at the surface of an
accreting NS can be divided into several different categories
according to their different sources of buoyancy, such as an entropy
gradient or density discontinuity. \cite{1995ApJ...449..800B} studied
surface g-modes in accreting systems with thermal buoyancy as the
restoring force. They obtained an analytic result for the mode
frequency in the non-rotating limit ($\Omega \rightarrow 0$)
\begin{align}
f_{th} &= 6.26 Hz \left(T_8\frac{16}{A}\right)^{\frac{1}{2}} \left(1+\left(\frac{3n\pi}{2\ln(\rho_b/\rho_t)}\right) ^2\right)^{-1/2}\nonumber\\ 
&\times\left(\frac{10 km}{R}\right) \left(\frac{l(l+1)}{2}\right)^{\frac{1}{2}},
\end{align}
where $T_8=\frac{T}{10^8 K}$, $A$ is the mass number, $l$ is the
spherical harmonic index, $n$ is the number of radial nodes in the
displacement eigenfunction, $R$ is the stellar radius, and $\rho_b$
and $\rho_t$ are densities at the bottom and top of the ocean,
respectively.  \cite{1996ApJ...467..773S} studied thermal g-modes in a
steady state accreting and nuclear burning atmosphere, and found some
modes can be excited by the $\epsilon$ mechanism (perturbations in the
nuclear burning). For example, see their Table 3 for results on the
oscillation periods of $g_1$ and $g_2$ modes.
\cite{1998ApJ...506..842B} studied the effect of hydrogen electron
captures on g-modes in the ocean of accreting neutron stars. They
found that the sudden increase in the density at the hydrogen electron
capture layer supports a density discontinuity mode with a
non-rotating limit frequency
\begin{equation}
\label{eq:f_d}
f_d \approx 35 Hz \left(\frac{X_r}{0.1}\right)^{\frac{1}{2}} \left(1-\frac{\Delta Z}{\Delta A}\right)^{\frac{1}{2}}\left(\frac{10 km}{R}\right) \left(\frac{l(l+1)}{2}\right)^{\frac{1}{2}},
\end{equation}
where $X_r$ is the residual mass fraction of hydrogen, and $\Delta Z$
and $\Delta A$ are the changes in the charge and mass of the nuclei
from one side of the discontinuity to the other.

\cite{2004ApJ...603..252P} studied non-radial oscillations at the
surface of a helium burning neutron star. They found one unstable mode
that resides in the helium atmosphere and is supported by the buoyancy
of the helium/carbon interface. The frequency of this mode in the
non-rotating limit is given by
\begin{equation}
f_{th-He}\approx (20-30 Hz) \sqrt{\frac{l(l+1)}{2}},
\end{equation}
which depends on the accretion rate. Similarly to the results of
\cite{1996ApJ...467..773S}, they find that this mode can also be
driven unstable by the $\epsilon$-mechanism, and they also compute
results for higher accretion rates.  Now, the orbital period of
J$1751$ is very short ($\sim 42$ min) \citep{2002ApJ...575L..21M},
which means that it is a very compact system and thus the donor star
is likely a helium white dwarf. This suggests that the accreted
material is helium-rich and it therefore seems plausible that the
system might show the unstable shallow surface waves that are obtained
for a helium atmosphere.

It is important to note that the g-mode frequencies just discussed are
obtained in the non-rotating limit $(\Omega \rightarrow 0)$ and,
as alluded to above, they will be modified by rapid rotation of the
star.  \cite{1996ApJ...460..827B} studied the effect of high spin
frequencies on the g-mode spectrum in the so-called ``traditional
approximation,'' in which mode propagation is confined to a thin
shell, the radial component of the coriolis force is neglected, and
the radial displacements produced by the modes are assumed to be 
much less than the horizontal displacements.  These approximations 
are reasonable for surface g-modes as long as the coriolis force remains 
less than the buoyant force (see \S 2 of \cite{1996ApJ...460..827B}). This condition
can be expressed as, $N^2 \gg (\Omega \omega R) / h$, where $N$, $R$
and $h$ are the Brunt V{\"a}is{\"a}ll{\"a} frequency (which sets the
strength of buoyancy), stellar radius, and characteristic scale height
in the surface envelope, respectively.  Assuming the candidate
frequency in J1751 represents the mode frequency in the co-rotating
frame (and using the 435 Hz spin frequency of J1751), then $\omega =
0.573 \Omega$, and we would require $N^2 \gg 4.3 \times 10^6 (R/h)$ Hz$^2$.

From this analysis \cite{1996ApJ...460..827B} found that stellar
rotation ``squeezes'' the eigenfunctions toward the equator within an
angle $\cos \theta < \frac{1}{q}$ where $\theta$ is measured from the
pole, $q=\frac{2 \Omega}{\omega}$, and the oscillation frequency of
the mode (in the co-rotating frame) in a non-rotating star,
$\omega_{l,0}$, is related to the mode frequency at arbitrary spin
frequencies by the following equation
 \begin{equation}
\omega^2=2\Omega \omega_{l,0}\left[\frac{(2l_{\mu}-1)^2}{l(l+1)}\right]^{1/2}
\end{equation}
where $l_{\mu}$ is the number of zero crossings in the angular
displacement between $\cos \theta =-\frac{1}{q}$ and $+\frac{1}{q}$.
Thus, the surface displacements of the rotationally modified g-modes
are strongly confined to the equatorial region at high spin
frequencies and the modes are exponentially damped for $\cos \theta
\geq 1/q$ \citep{1996ApJ...460..827B}. For the spin frequency of
J1751, $q \simeq 3.5$ (assuming that the 249.33 Hz candidate frequency
is associated with a co-rotating frame mode frequency). This leaves
open the question of what mode amplitudes would be needed to
effectively perturb a hot-spot located near the rotational pole.

Moreover, the relevant scale height, $h$, will depend on details
of the surface envelopes in question, however, a typical value for the
He-rich envelopes of Piro \& Bildsten (2004) is $h \approx 200$ cm, thus, 
for a 10 km radius neutron star we
would require $N^2 \gg 2.2 \times 10^{10}$ Hz$^2$ in order to satisfy
the assumptions associated with the traditional approximation. We note
that this condition appears to be technically violated for these
envelopes as $N$ is everywhere less than about $1 \times 10^5$ Hz (see
their Figures 2 and 3).  This suggests the need for more theoretical
work in order to more accurately determine the surface g-mode properties
for rotation rates appropriate to the faster spinning AMXPs (such as
J1751). In addition, more work similar to that done in the context of
r-modes by \cite{2010MNRAS.409..481N} should be done for the
rotationally modified g-modes to determine how efficiently these modes
can perturb a hot-spot located near the spin axis of the star and the
resulting light curves.

Keeping in mind these caveats, we can nevertheless rearrange Eq. 4
to express the frequency of the mode in a non-rotating star,
$f_{l,0}$, in terms of the observed mode co-rotating frame frequency,
$f_{obs}$, stellar spin frequency, $\nu_{spin}$ and the mode indices
$l$, $m$ and $l_{\mu}$.  If the observed frequency is directly related
to the modes co-rotating frame frequency (the ``co-rotating frame
scenario''), we find,
\begin{equation}
f_{l,0}=(f_{obs}^2/(2\nu_{spin}))*\sqrt{(l(l+1)/(2l_{\mu}-1)^2} \; .
\end{equation}
However, if the observed oscillation frequency is the modes inertial
frame frequency (the ``inertial frame scenario''), then we must first
relate this to the co-rotating frame via $f_{obs} = m\nu_{spin} -
f_{obs, i}$, where $f_{obs, i}$ is the (observed) inertial frame mode
frequency, yielding,
\begin{equation}
f_{l,0}=( (m\nu_{spin} - f_{obs, i})^2/(2\nu_{spin}))*\sqrt{(l(l+1)/(2l_{\mu}-1)^2} \; .
\end{equation}
We can then find plausible non-rotating g-modes that can be consistent
with the candidate frequency. Possible identifications are summarized
in Table 1 and discussed in more detail below.

From the discussion above we can see that the candidate peak at
$0.5727\times \nu_{spin} = 249.33$ Hz in J$1751$ may be identified as
an $l=2,\; m=1 \; (l_{\mu}=3)$ g-mode that resides in a helium
atmosphere and has a non-rotating frequency of $\sim 35$ Hz as
observed in the co-rotating frame. This is based on the assumption
that this surface mode perturbs the hot-spot periodically and the
candidate frequency is related to the frequency of the mode in the
co-rotating frame. This mode is consistent with the $g_2$ mode given
in Table 3 of \cite{1996ApJ...467..773S} with a period of 29.04 ms and
$\dot{M}/\dot{M}_{Edd}=0.7$ in a pure helium shell. The thermal g-mode
computations discussed above have been done under the assumption of
steady-state nuclear burning in a thermally stable envelope. Now,
stable burning of the accreted material in the envelope requires a
high, near-Eddington accretion rate \citep{2004ApJ...603..252P},
however, the average accretion rate of J1751 was about $2.1 \times
10^{-11} M_{\odot}$ yr$^{-1}$ \citep{2002ApJ...575L..21M}, which is
low relative to the Eddington rate, and therefore the assumption of
steady-state nuclear burning may not be applicable in this
case. However, we note that the relevant accretion rate for the
thermal stability calculation is the local value (per unit area) which
might be higher depending on the accretion geometry, for example, if
accretion is restricted to a portion of the neutron star's surface.

If we assume that the amplitude of this mode is high enough that it
can modify the temperature distribution at the surface of the star and
produce observable X-ray variations then the inertial frame scenario
is relevant (see Eq. 6 above).
In  this case the candidate oscillation in J1751 may be consistent with an
$l=m=1$ shallow surface wave in the helium layer with a non-rotating
frequency of 18.7 Hz (this is slightly less than the lower limit of
20 Hz given in \cite{2004ApJ...603..252P}).

Another possibility for the candidate at $249.33$ Hz would
be an $l=l_{\mu}=2$ (with $m=0$ or $m=2$) density discontinuity g-mode
due to hydrogen electron captures in the ocean of the star with a
non-rotating limit frequency of $f_d \simeq 58.34$ (see
Eq.~\ref{eq:f_d}) as measured in the co-rotating frame. However,
whether or not sufficient hydrogen is present to support a density
discontinuity mode in such a compact and presumably helium-rich system
as J1751 remains an open question.

\cite{1986ApJ...305..767C} showed that in the presence of strong
magnetic fields the frequencies and displacements of modes that reside
in the ocean, in particular g-modes, will be modified. For magnetic
fields $B_0 > 10^5$ G, these modified g-modes (magneto-gravity modes)
change with increasing $B$ from predominantly g-modes with constant
periods to predominantly magnetic modes with periods proportional to
$B_0^{-1}$ (see their Eq. 42 and Figure 4).
\cite{2004ApJ...603..252P} estimated the maximum magnetic field before
the shallow surface mode would be dynamically affected to be $B_{dyn}
\approx 5 \times 10^7 \rm{G} (\frac{\omega/2 \pi}{21.4 Hz})$ which is
about $6\times 10^8$ G for a rotationally modified shallow surface
wave with a frequency of 249.33 Hz. This is close to the estimated
value of the magnetic field of J1751 obtained from spin-down
measurements due to magneto-dipole radiation which is about $4 \times
10^8$ G \citep{2011A&A...531A.140R}.  However, \cite{2009ApJ...703.1819H} 
also explored the effect of a vertical magnetic field on
shallow surface waves, and their results suggest that for the spin
rate and likely magnetic field strength appropriate to J175, the field
does not strongly modify the mode frequencies (see the
``magneto-Poincare modes'' in their Figure 2).  The above results
support the conclusion that the magnetic field likely does not exert a
dramatic influence on these g-mode frequencies.

\subsection{r-modes and inertial modes}
As we discussed earlier, another class of non-radial oscillation modes
that may have frequencies consistent with the candidate signal in
J$1751$ are the r-modes. A 3-component neutron star model may have
unstable r-modes in the ocean and/or in the core. The frequency of the
r-modes in the slow-rotation limit ($\bar{\Omega} \equiv
\Omega/(GM/R^3)^{1/2} \rightarrow 0$) is given by
$\omega_0=2m\Omega/[l(l+1)]$.  As the rotation frequency of the star
increases, the co-rotating frame frequency of the r-modes in the
surface layer of the star deviates appreciably from this asymptotic
form and becomes almost insensitive to $\Omega$ (see Figure 4 in
\cite{2004ApJ...600..914L}). According to Table 2 in
\cite{2004ApJ...600..914L} the frequencies of the surface r-modes are
always less than 200 Hz for the spin frequencies and mass accretion
rates that are relevant to LMXBs. For example, for the $l=|m|=2$
fundamental r-modes of radiative envelopes \cite{2004ApJ...600..914L}
found that the co-rotating frequency is in the range of 101 to 173 Hz
for the stellar spin frequencies of 300 to 600 Hz and $\dot{M}=0.02
\dot{M}_{Edd}$ to $0.1 \dot{M}_{Edd}$. \cite{2010MNRAS.405.1444L} also
studied the low frequency oscillations of rotating and magnetized
neutron stars and found no r-modes confined in the ocean in the
presence of a magnetic field even as low as $B_0 \sim 10^7$ G in a 3
component NS model. Thus, the candidate frequency at 249.33 Hz in
J$1751$ doesn't appear to be consistent with that of a surface r-mode.

Although the amplitudes of the ocean g- and r-modes tend to be
confined to the equatorial regions, this is not the case for $l = |m|$
r-modes in the fluid core. In fact \cite{2010MNRAS.405.1444L} showed
that the displacement vector of these core r-modes have large
amplitudes around the rotation axis at the stellar surface even in the
presence of a surface magnetic field $B_0 \sim 10^{10}$ G.

As we briefly mentioned in the previous sections, the co-rotating
frame frequency of $l=m=2$ core r-modes (which are the most unstable
ones) in the $\bar{\Omega} \rightarrow 0$ limit is equal to
$\omega_0=\frac{2}{3}\Omega$ which is larger than the frequency of the
candidate peak at $\omega=0.5727 \Omega$ and adding the corrections
due to high spin rates only slightly increases the slow-rotation limit
value. \cite{2001ApJ...546.1121Y} studied the effect of a solid crust
on the r-mode oscillations of a three component NS model.  At
sufficiently small values of $\Omega$ the coupling between r-modes and
crustal toroidal modes is negligibly weak, but at higher spin
frequencies they found that the core r-modes are strongly affected by
the mode coupling with crustal toroidal modes, and because of the
avoided crossings with the crustal toroidal modes, the core r-modes
will lose their simple form of eigenfrequency and eigenfunction.  The
r-mode frequency increases as the spin frequency of the star
increases, and at some point it meets the frequency of the crustal
modes which results in avoided crossings. Depending on the thickness
of the crust and therefore the number of modes in the crust with
relevant frequencies, there might be several avoided crossings between
core r-modes and crustal toroidal modes \citep{2001MNRAS.324..917L,
2006PhRvD..74d4040G}. The spin frequencies at which the avoided
crossings occur are given by $ \Omega_{cross} \approx
\frac{l(l+1)}{m}\omega_t(0)$ where $\omega_t(0)$ is the oscillation
frequency of the toroidal mode at $\Omega=0$ and it is a function of
the shear modulus of the crust.  As shown in Figure 3 of
\citet{2001ApJ...546.1121Y}, in the presence of a solid crust and at
high rotation frequencies, the r-mode frequency in the co-rotating
frame deviates from its simple form in the $\bar{\Omega} \rightarrow
0$ limit.
For fundamental r-modes with $l=m=2$ they showed that $\kappa$ can
decrease from its slow-rotation limit and span a range of values from
$\frac{2}{3}$ to less than $0.4$ depending on the spin frequency of
the star and the properties of the solid crust, such as its shear
modulus. We note that the value of $\mu / \rho$ is almost constant in
the crust of a neutron star, $\mu / \rho \simeq 1-6 \times 10^{16}$
cm$^2$ s$^{-2}$ (see for example Figure 1 in
\cite{2006PhRvD..74d4040G}). The results of \cite{2001ApJ...546.1121Y}
given in their Figure 3 and Table 1 suggest that for $\kappa \sim
0.57$ at $\bar{\Omega} \simeq 0.2$ (relevant for J1751) one needs the
shear modulus of the crust to be a few times higher than the standard
values given by \cite{1991ApJ...375..679S} for a {\it bcc} crystal at
the higher densities in the crust. This suggests that observations of
r-mode induced oscillations in the X-ray flux of neutron stars could
be useful in probing the structure and properties of the crust.

In addition to the r-modes the Coriolis force also supports the
more general class of inertial modes which have both significant
toroidal and spheroidal angular displacements, whereas the r-modes are
principally toroidal.  A number of authors have studied the properties
of inertial modes, and in particular their relationship to other
low-frequency modes such as the g-modes \citep{2000ApJ...529..997Y,
2000ApJS..129..353Y, 2009MNRAS.394..730P, 2010MNRAS.405.1444L}.  For
example, \cite{2009MNRAS.394..730P} have computed time evolutions of
the linear perturbation equations in order to explore the oscillations
of rapidly rotating, stratified (non-isentropic) neutron stars. They
find that the g-modes in stratified stars become strongly modified by
rapid rotation, with each g-mode frequency approaching that of a
particular inertial mode associated with the corresponding isentropic
(ie. no bouyancy) stellar model.  Earlier work by
\cite{2000ApJS..129..353Y} reached a similar conclusion, but the more
recent results of \cite{2009MNRAS.394..730P} have explored the
connection to much higher rotation rates. These studies, as well as
the recent calculations of \cite{2010MNRAS.405.1444L}, all find some
inertial modes with co-rotating frame frequencies that appear at least
qualitatively consistent with the candidate oscillation in J1751.  For
example, the $^3i_1$ and $^4i_2$ modes of \cite{2009MNRAS.394..730P}
have frequencies near $\omega = 0.573\Omega$ (see their Table 2 and
Figures 3 and 11).  Note that for their stellar models $\Omega / (G
\rho_c)^{1/2} \approx 0.5$ is appropriate for the 435 Hz spin
frequency of J1751.  Similarly, the $l_0 - |m| = 2$, $m=2$ prograde,
isentropic inertial mode of \cite{2000ApJ...529..997Y}, and the
non-isentropic modes labelled $g_{-1}(2) <-->i_{-1}(2)$ in Figure 9a of
\cite{2000ApJS..129..353Y} have frequencies near
to that of our candidate oscillation. It should be noted, however,
that all these calculations have significant simplifications that
likely make detailed quantitative comparisons with our observed
frequency problematic.  For example, they all employ rather
simplistic stellar models, such as the use of polytropic equations of
state, and the models do not have a solid crust.  Additionally, the
calculations of Yoshida \& Lee (2000a,b) were for relatively modest
spin rates, and extrapolation to the higher spin rate appropriate for
J1751 is perhaps risky.

In addition, \cite{2010MNRAS.405.1444L} has presented oscillation mode
calculations for rotating, and magnetized neutron stars using
3-component (ocean, crust, core) models.  He also finds prograde
inertial modes with frequencies approximately consistent with our
candidate oscillation (see, for example, the $|m|=2$ modes for $B_0 =
10^{10}$ G near the lower right corner of Figure 5).  These
calculations were for a low mass, $0.5 M_{\odot}$, neutron star and
are also only strictly valid for modest rotation rates, so, again,
caution should be exercised when making quantitative comparisons with
observed frequencies.  We emphasize that all of the above calculations
were for global stellar modes, and not restricted to only surface
displacements. Similarly to the global r-modes these inertial modes
will likely have appreciable surface amplitudes closer to the
rotational poles than the surface-based, rotationally modified g-modes
investigated by \cite{1996ApJ...460..827B} and
\cite{2004ApJ...603..252P}.  Based on the above discussion it seems
possible that inertial modes could be relevant to our candidate
oscillation in J1751, but clearly new theoretical work is needed to
explore such modes in more realistic, rapidly rotating neutron star
models before any firm conclusion should be drawn. Further, new
calculations to determine how effectively inertial modes can perturb
an X-ray emitting hot-spot, and the resulting light curves, are
certainly warranted.

\subsection{Coherence of the Candidate Oscillation}

The candidate power spectral peak in J$1751$ is narrow, which means
that the oscillation frequency has to be steady over most of the time
span used to compute the power spectrum, which is about 6 days. Thus,
if the candidate peak is due to some non-radial oscillation of the
star, its frequency has to be almost constant during that time
span. Between surface g-modes and core r-modes which might be
consistent with the observed candidate peak as discussed above,
r-modes are expected to have steady frequencies over such a short time
span because they reside in the core and conditions there are not
expected to change over such timescales. Among surface g-modes that
are consistent with the candidate oscillation in J$1751$, thermal
g-modes of a helium burning neutron star reside in the shallow layers
close to the surface of the star, but the density discontinuity
g-modes due to hydrogen electron capture reside in deeper layers close
to the ocean-crust interface.  If the temperature and elemental
composition of the ocean doesn't change during the time span used to
compute the light curve, the frequency would be steady which is
expected to be the case if the accretion rate varies little. In fact,
it has been shown by \cite{2004ApJ...603..252P} that the g-mode
frequency scales approximately as $\dot{m}^{1/8}$ where $\dot{m}$ is
the local accretion rate, and therefore a small change in the
accretion rate will not have a large effect on the g-mode frequencies.

The light curve of J1751 (see Figure~\ref{fig:J1751-lc}) shows
variation in the count rate at the level of 30-40 counts s$^{-1}$,
which likely suggests some variation in the accretion rate. Although
we note that X-ray flux (or count rate) is known to not always
correlate linearly with the accretion rate.  While this suggests the
mode frequency may change, a second effect likely limits the rate at
which it can vary, and that is set by the time, $t_{acc}$, required to
change conditions in the surface layers at a column depth where the
mode frequency is set.  This can be roughly approximated as $t_{acc}
\approx y / \dot m$, where $y$ is the relevant column depth in g
cm$^{-2}$, and $\dot m$ is the accretion rate per unit surface area.
For an accretion rate of $2 \times 10^{-11} M_{\odot}$ yr$^{-1}$, and
a characteristic column depth of $10^{8}$ g cm$^{-2}$, $t_{acc}$ is
about 11.6 days. So, while accretion rate variations can, in
principle, change the g-mode frequencies, for timescales much less
than $t_{acc}$ the frequency is likely reasonably stable.

\subsection{Future Capabilities and Sensitivities}

As can be seen in several of our power spectra (see for example,
Figure 7), an upper limit on the modulation amplitude is approximately
given by $1/(N_{tot})^{1/2}$, where $N_{tot}$ is simply the total
number of X-ray events in the light curve from which the power
spectrum is computed.  The approximation is better as one averages
more frequency bins, meaning it is a more precise limit in the context
of broader band-width signals. For the full resolution spectra
presented here the derived upper limits are reasonably approximated as
$\approx 4/(N_{tot})^{1/2}$.

This is not too surprising, as the fractional Poisson error on the
average count rate within a time interval is just
$1/(N_{tot})^{1/2}$. Thus, this limit is simply a statement that one
cannot measure a fractional modulation amplitude of the X-ray count
rate that is smaller than the precision with which that rate can be
determined.  Assuming that other necessary capabilities are present in
future observatories---such as adequate high frequency time
resolution---then a simple way to estimate the amplitude sensitivity
for future detectors is just to scale up the expected count rates
appropriately. The above considerations are valid in the case that the
source count rate dominates any background rate.

The largest effective area for fast X-ray timing presently being
planned is ESA's Large Observatory for X-ray Timing ({\it LOFT},
Feroci et al. 2012). The Large Area Detector (LAD) on {\it LOFT} would
consist of $\approx 12$ m$^2$ of silicon detectors and due to the
larger collecting area and better (flatter) response above 6-7 keV
would provide an increase in source count rate compared to the PCA on
{\it RXTE} of about a factor of 30 (though the exact scaling would
depend on the X-ray spectrum of the source being considered).  The
other way to increase the total counts that can be included in a light
curve is to more densely sample an outburst, and to Fourier analyse
longer continuous time intervals. For the sake of argument, if we
scale based on the most sensitive observation reported here, that is,
the single $\approx 6$ day interval for J1751, and assume that a {\it
LOFT} observation has twice the duty cycle and extends for twice as
long, then we might expect to reach an amplitude limit of $a_{amp}
\approx 1/(2*2*30*44\times 10^6)^{1/2} = 1.4 \times 10^{-5}$.

While this represents a limit on the Fourier amplitude of X-ray flux
modulations that could be detected, the corresponding amplitude of an
oscillation mode would depend on the details of how the oscillation
mode perturbs the X-ray emission.  \cite{2010MNRAS.409..481N} show
from their light curve modeling that the observed Fourier amplitude is
proportional to the normalized amplitude of the stellar oscillation
(see their Figure 6). The details of the scaling depends on the
particular oscillation mode and other details, but a rough estimate
indicates that the Fourier amplitude $a_{amp} \approx 1-2 \times A$,
where $A$ represents the maximum horizontal displacement produced by a
mode divided by the stellar radius ($A = {\rm
max}(|\xi_{\theta}|/R,|\xi_{\phi}|/R)$). Based on this simple scaling
one can expect that future sensitivities with {\it LOFT} would be such
that $A \approx 1\times 10^{-5}$ could be probed.  We note that this
corresponds to a 10 cm maximum surface displacement for a 10 km
neutron star.

In the case of r-mode oscillations, $A$ is approximately equal to $\alpha/2$, where $\alpha$ is the
dimensionless amplitude of the mode, defined in Eq. 1 of
\cite{1998PhRvL..80.4843L}. We note that for the candidate oscillation
in J$1751$, $A \approx 7\times 10^{-4}$, and $\alpha \sim 10^{-3}$. This is much larger than the
upper limits on $\alpha$ given in \cite{2013ApJ...773..140M}, which is
less than $10^{-7}$ for J1751 (see also
\cite{2012MNRAS.424...93H}).  A global r-mode with an amplitude of the order of $10^{-3}$ would cause a rapid spin-down of the star. Using the corresponding equation for spin-down due to gravitational wave emission from unstable r-modes \citep{1998PhRvD..58h4020O}, $d\Omega/dt\simeq-(2\Omega/\tau_V)Q\alpha^2$, where $\tau_V(T,\Omega,\alpha)$ is the viscous damping timescale of the mode, and $Q \equiv \frac{3 \tilde{J}}{2 \tilde{I}}$ \citep{2013ApJ...773..140M}, gives a spin-down rate of $\sim -1.3\times 10^{-9}$ Hz s$^{-1}$ for J1751 assuming a core temperature of $\sim 3\times 10^7 K$ \citep{2013ApJ...773..140M}. We note that even with a higher core temperature of $\sim 3\times 10^8 K$ the spin-down rate would be $\sim -2.8\times 10^{-11}$ Hz s$^{-1}$, which would still dominate the accretion spin-up rate and therefore is inconsistent with the observations \citep{2012arXiv1206.2727P}.  Further, if the amplitude of the mode is saturated at $\alpha_s \sim 10^{-3}$, $\tau_V$ in the spin evolution equation should be replaced by $\tau_G$, where $\tau_G$ is the gravitational radiation timescale. This would cause an even larger spin-down rate of $\approx - 1.5 \times 10^{-7}$ Hz s$^{-1}$.  Such a large amplitude for a global r-mode, even assuming that it is large only during the outburst and would be damped in quiescence, would cause a large change in the frequency of J1751 which would be easily detectable in the data. In addition, the maximum saturation amplitude due to nonlinear mode coupling, computed by \cite{2003ApJ...591.1129A}, $\alpha_s\approx 8\times 10^{-3} (\nu_s/1 kHz)^{5/2}$, that in the case of J1751 is $\sim 6\times10^{-5}$ (see also \cite{2008MNRAS.389..839W} and \cite{2007PhRvD..76f4019B}), and the upper limits on $\alpha$ ($\sim 10^{-4}$) from gravitational wave searches with LIGO \citep{2010PhRvD..82j4002O,2013arXiv1309.4027A} further support the notion that the candidate oscillation is unlikely to be a global r-mode. This argues that a g-mode or inertial mode
interpretation is more likely. While we think the present evidence is
strongly suggestive, future, more sensitive observations will likely
be needed to confirm the presence of non-radial oscillation modes in
J$1751$ and/or other AMXPs.

\section{Summary and Conclusions}

We have carried out searches for X-ray modulations that could be
produced by global non-radial oscillation modes in several AMXPs.  A
likely mechanism for generating X-ray flux modulations is that due to
perturbations to the X-ray emitting hot-spot produced by surface
motions associated with the oscillation modes (see for example, Numata
\& Lee 2010).  In this regard the most relevant non-radial modes are
those with predominantly horizontal displacements at the stellar
surface, such as the inertial modes (which includes the r-modes), 
and the g-modes. In order to search most sensitively for nearly coherent 
modulations we first remove the Doppler delays due to the binary motion of 
the neutron star.  We search a range of frequencies--scaled to the stellar spin
frequency--that are theoretically consistent with those expected for
the global r-modes in neutron stars, and this range also encompasses
the frequencies expected for some surface g-modes.  We find one
plausible candidate signal in J1751 with an estimated significance of
$1.6 \times 10^{-3}$, and upper limits for the two other sources we
studied, X-2, and J1814.

Our candidate signal in J1751 appears at a frequency of
$0.5727597\times \nu_{spin} = 249.332609$ Hz, has a fractional Fourier
amplitude of $7.455 \times 10^{-4}$, and is effectively coherent over
the entire light curve in which it was found. Based on its observed
frequency it appears at least plausible that it could be related to a
surface g-mode associated with a helium-rich layer on the neutron star
surface \citep{2004ApJ...603..252P}.  Other possibilities include a
g-mode associated with density discontinuities in the surface layers
\citep{1998ApJ...506..842B}, an inertial mode \citep{2009MNRAS.394..730P}, or 
perhaps an r-mode modified by the presence of the neutron star crust 
\citep{2001ApJ...546.1121Y}.

For J1814 we find an amplitude upper limit to any signal of $\approx
7.8\times 10^{-4}$ (for a coherent signal). For broader bandwidth
signals the limit approaches $\approx 2.2 \times 10^{-4}$. In the case
of X-2, because less data is available for this source, the limits are
less constraining, and we find values of $5.6 \times 10^{-3}$ and $2.8
\times 10^{-3}$ at frequency resolutions of $3.125 \times 10^{-4}$ and
$0.01$ Hz, respectively.

\acknowledgments We thank Tony Piro, Andrew Cumming, Jean in 't Zand,
Cole Miller, and Diego Altamirano for many helpful comments and
discussions. We thank the anonymous referee for valuable comments that helped us
improve this paper.  TS acknowledges NASA's support for high energy
astrophysics. SM acknowledges the support of the U.S. Department of
Energy through grant number DEFG02- 93ER-40762.

\bibliography{ms_rev21}
\newpage

\begin{table*}
\renewcommand{\arraystretch}{1.5}
\caption{ Possible {\it\lowercase{g}}-mode identifications}
\scalebox{1}{
\begin{tabular}{cccccccc}
\tableline\tableline
 & $l$ &  $m$& $l_{\mu}$ &  $\omega_{l,0} /(2\pi)$& Consistent with &$\omega_{l,0} /(2\pi)$& Consistent with \tabularnewline
&& && rotating frame &&inertial frame & \tabularnewline
&& && scenario  &&scenario & \tabularnewline
\tableline\tableline
& & & & & & & \\
 
&1& 0&1 & $101.1$ & &101.1& \\

&1& 1 &2 & $33.7$ &Thermal g-mode (helium atmosphere) &18.7&Thermal g-mode (helium atmosphere)\\

&2&-1 & $\sim1$ &  $175.0$&&1318.5&\\

&2&0 &2 &  $58.3$&Density discontinuity g-mode&58.3&Density discontinuity g-mode\\

&2&1 &3 &  $35.0$&Thermal g-mode (helium atmosphere)&19.4&\\
&& & & &Density discontinuity g-mode&&\\

&2&2 &2 &  $58.3$&Density discontinuity g-mode&361.5&\\
\tableline
\end{tabular}}
\tablecomments{Potential g-mode identifications for the candidate
oscillation frequency in J1751. Values of $l$, $m$, $l_{\mu}$, and the
corresponding non-rotating mode frequency, $\omega_{l,0}$, are given
for both the case in which the observed frequency is a rotating or
inertial frame frequency.  The column labelled ``Consistent with''
gives the specific type of g-mode/s that could be consistent with the
given frequencies.  No entry in this column indicates that the
indicated frequencies are unlikely for any of the specific g-modes
discussed in \S 4. For completeness we note that for the even parity
g-modes with $m=-1$ and $m=-2$, the effective wavenumber, $\lambda$,
goes to $m^2$ for high spin rates, which wouldn't be consistent with
the candidate frequency \citep{2004ApJ...603..252P}. Finally,
the existence of the density discontinuity modes requires that
hydrogen is present in the surface layers at depths sufficient to
initiate electron captures \citep{1998ApJ...506..842B}. }
\end{table*}

\newpage

\end{document}